\documentclass[numberedappendix]{emulateapj} 
\pdfoutput=1
\usepackage{amsmath,natbib,graphicx}
\usepackage{epsf}
\input{epsf}
\usepackage{epstopdf}
\usepackage{epsfig}
\usepackage{rotating}
\usepackage{fmtcount}
\usepackage{appendix}
\DeclareGraphicsExtensions{.jpg,.pdf,.png,.eps,.ps}
\graphicspath{{FIGURES/}}

\newcommand{\LCDM}{\mbox{$\Lambda$CDM}}
\newcommand{\wCDM}{\mbox{wCDM}}
\newcommand{\rcore}{\mbox{$r_\mathrm{core}$}}

\newcommand{\ltsima}{$\; \buildrel < \over \sim \;$}
\newcommand{\ltsim}{\lower.5ex\hbox{\ltsima}}
\newcommand{\onefifty}{$150\,$GHz}
\newcommand{\fiveh}{$5^\mathrm{h}$}
\newcommand{\twentythreeh}{$23^\mathrm{h}$}

\newcommand{\sqdeg}{\ensuremath{\mathrm{deg}^2}}
\newcommand{\uk}{\mbox{$\mu \mbox{K}$}}
\newcommand{\msun}{\ensuremath{M_\odot}}
\newcommand{\SN}{\ensuremath{\xi}}
\newcommand{\nSN}{\ensuremath{\zeta}}
\newcommand{\tabd}{\ensuremath{\dagger}}
\newcommand{\tabdd}{\ensuremath{\ddagger}}

\def\McGill{1}
\def\KICPChicago{2}
\def\AAUChicago{3}
\def\Yale{4}
\def\Cardiff{5}
\def\UChicago{6}
\def\Berkeley{7}
\def\EFIChicago{8}
\def\PhysicsUChicago{9}
\def\CfA{10}
\def\Illinois{11}
\def\Davis{12}
\def\Colorado{13}
\def\Harvard{14}
\def\NASA{15}
\def\LBNL{16}
\def\Michigan{17}
\def\Munich{18}
\def\ExcellenceCluster{19}
\def\MPE{20}
\def\CaseWestern{21}
\def\Taiwan{22}

\begin{document}

\title{Galaxy Clusters Selected with the Sunyaev-Zel'dovich Effect \\
	from 2008 South Pole Telescope Observations}

%\slugcomment{Submitted to \apj}

\author{
K.~Vanderlinde,\altaffilmark{\McGill}
T.~M.~Crawford,\altaffilmark{\KICPChicago,\AAUChicago}
T.~de~Haan,\altaffilmark{\McGill}
J.~P.~Dudley,\altaffilmark{\McGill}
L.~Shaw,\altaffilmark{\McGill,\Yale}
P.~A.~R.~Ade,\altaffilmark{\Cardiff}
K.~A.~Aird,\altaffilmark{\UChicago}
B.~A.~Benson,\altaffilmark{\Berkeley,\KICPChicago,\EFIChicago}
L.~E.~Bleem,\altaffilmark{\KICPChicago,\PhysicsUChicago}
M.~Brodwin,\altaffilmark{\CfA}
J.~E.~Carlstrom,\altaffilmark{\KICPChicago,\AAUChicago,\EFIChicago,\PhysicsUChicago} 
C.~L.~Chang,\altaffilmark{\KICPChicago,\EFIChicago}
A.~T.~Crites,\altaffilmark{\KICPChicago,\AAUChicago}
S.~Desai,\altaffilmark{\Illinois}
M.~A.~Dobbs,\altaffilmark{\McGill}
R.~J.~Foley,\altaffilmark{\CfA} 
E.~M.~George,\altaffilmark{\Berkeley}
M.~D.~Gladders,\altaffilmark{\KICPChicago,\AAUChicago}
N.~R.~Hall,\altaffilmark{\Davis}
N.~W.~Halverson,\altaffilmark{\Colorado}
F.~W.~High,\altaffilmark{\Harvard} 
G.~P.~Holder,\altaffilmark{\McGill}
W.~L.~Holzapfel,\altaffilmark{\Berkeley}
J.~D.~Hrubes,\altaffilmark{\UChicago}
M.~Joy,\altaffilmark{\NASA}
R.~Keisler,\altaffilmark{\KICPChicago,\PhysicsUChicago}
L.~Knox,\altaffilmark{\Davis}
A.~T.~Lee,\altaffilmark{\Berkeley,\LBNL}
E.~M.~Leitch,\altaffilmark{\KICPChicago,\AAUChicago}
A.~Loehr,\altaffilmark{\CfA} 
M.~Lueker,\altaffilmark{\Berkeley}
D.~P.~Marrone,\altaffilmark{\KICPChicago,\UChicago}
J.~J.~McMahon,\altaffilmark{\KICPChicago,\EFIChicago,\Michigan}
J.~Mehl,\altaffilmark{\KICPChicago,\AAUChicago}
S.~S.~Meyer,\altaffilmark{\KICPChicago,\EFIChicago,\PhysicsUChicago,\AAUChicago}
J.~J.~Mohr,\altaffilmark{\Munich,\ExcellenceCluster,\MPE}
T.~E.~Montroy,\altaffilmark{\CaseWestern}
C.-C.~Ngeow,\altaffilmark{\Illinois,\Taiwan}
S.~Padin,\altaffilmark{\KICPChicago,\AAUChicago}
T.~Plagge,\altaffilmark{\Berkeley,\AAUChicago}
C.~Pryke,\altaffilmark{\KICPChicago,\AAUChicago,\EFIChicago} 
C.~L.~Reichardt,\altaffilmark{\Berkeley}
A.~Rest,\altaffilmark{\Harvard}
J.~Ruel,\altaffilmark{\Harvard}
J.~E.~Ruhl,\altaffilmark{\CaseWestern} 
K.~K.~Schaffer,\altaffilmark{\KICPChicago,\EFIChicago} 
E.~Shirokoff,\altaffilmark{\Berkeley} 
J.~Song,\altaffilmark{\Illinois}
H.~G.~Spieler,\altaffilmark{\LBNL}
B.~Stalder,\altaffilmark{\CfA}
Z.~Staniszewski,\altaffilmark{\CaseWestern}
A.~A.~Stark,\altaffilmark{\CfA} 
C.~W.~Stubbs,\altaffilmark{\Harvard,\CfA} 
A.~van~Engelen,\altaffilmark{\McGill}
 J.~D.~Vieira,\altaffilmark{\KICPChicago,\PhysicsUChicago}
R.~Williamson,\altaffilmark{\KICPChicago,\AAUChicago} 
Y.~Yang,\altaffilmark{\Illinois}
O.~Zahn,\altaffilmark{\Berkeley}
and
A.~Zenteno\altaffilmark{\Munich,\ExcellenceCluster}
}

\altaffiltext{\McGill}{Department of Physics, McGill University, 3600 Rue University, Montreal, Quebec H3A 2T8, Canada}
\altaffiltext{\KICPChicago}{Kavli Institute for Cosmological Physics, University of Chicago, 5640 South Ellis Avenue, Chicago, IL 60637}
\altaffiltext{\AAUChicago}{Department of Astronomy and Astrophysics, University of Chicago, 5640 South Ellis Avenue, Chicago, IL 60637}
\altaffiltext{\Yale}{Department of Physics, Yale University, P.O. Box 208210, New Haven, CT 06520-8120}
\altaffiltext{\Cardiff}{Department of Physics and Astronomy, Cardiff University, CF24 3YB, UK}
\altaffiltext{\UChicago}{University of Chicago, 5640 South Ellis Avenue, Chicago, IL 60637}
\altaffiltext{\Berkeley}{Department of Physics, University of California, Berkeley, CA 94720}
\altaffiltext{\EFIChicago}{Enrico Fermi Institute, University of Chicago, 5640 South Ellis Avenue, Chicago, IL 60637}
\altaffiltext{\PhysicsUChicago}{Department of Physics, University of Chicago, 5640 South Ellis Avenue, Chicago, IL 60637}
\altaffiltext{\CfA}{Harvard-Smithsonian Center for Astrophysics, 60 Garden Street, Cambridge, MA 02138}
\altaffiltext{\Illinois}{Department of Astronomy, University of Illinois, 1002 West Green Street, Urbana, IL 61801}
\altaffiltext{\Davis}{Department of Physics, University of California, One Shields Avenue, Davis, CA 95616}
\altaffiltext{\Colorado}{Department of Astrophysical and Planetary Sciences and Department of Physics, University of Colorado, Boulder, CO 80309}
\altaffiltext{\Harvard}{Department of Physics, Harvard University, 17 Oxford Street, Cambridge, MA 02138}
\altaffiltext{\NASA}{Department of Space Science, VP62, NASA Marshall Space Flight Center, Huntsville, AL 35812}
\altaffiltext{\LBNL}{Physics Division, Lawrence Berkeley National Laboratory, Berkeley, CA 94720}
\altaffiltext{\Michigan}{Department of Physics, University of Michigan, 450 Church Street, Ann Arbor, MI, 48109}

\altaffiltext{\Munich}{Department of Physics, Ludwig-Maximilians-Universit\"{a}t, Scheinerstr.\ 1, 81679 M\"{u}nchen, Germany}
\altaffiltext{\ExcellenceCluster}{Excellence Cluster Universe, Boltzmannstr.\ 2, 85748 Garching, Germany}
\altaffiltext{\MPE}{Max-Planck-Institut f\"{u}r extraterrestrische Physik, Giessenbachstr.\ 85748 Garching, Germany}
\altaffiltext{\CaseWestern}{Physics Department and CERCA, Case Western Reserve University, 10900 Euclid Ave., Cleveland, OH 44106}
\altaffiltext{\Taiwan}{Graduate Institute of Astronomy, National Central University, No. 300 Jonghda Rd, Jhongli City, 32001, Taiwan}

\email{keith.vanderlinde@mail.mcgill.ca}

\begin{abstract}

We present a detection-significance-limited catalog of 21 Sunyaev-Zel'dovich-selected galaxy clusters. 
These clusters, along with 1 unconfirmed candidate, were identified in
178 deg$^2$ of sky surveyed in 2008 by the South Pole Telescope 
to a depth of 18 $\mu$K-arcmin at $150\,$GHz. 
Optical imaging from the Blanco Cosmology Survey (BCS) and Magellan telescopes provided
photometric (and in some cases spectroscopic) redshift estimates, 
with catalog redshifts ranging from $z=0.15$ to $z>1$, with a median $z = 0.74$. 
Of the 21 confirmed galaxy clusters, 
three were previously identified as Abell clusters, 
three were presented as SPT discoveries in \citet{staniszewski09}, 
and three were first identified in a recent analysis of BCS data by
\citet{menanteau10}; the remaining 12 clusters are presented for the first time in this work.
Simulated observations of the SPT fields predict the sample
to be nearly 100\% complete above a mass threshold of
$M_{200} \approx 5\times10^{14}\,\msun h^{-1}$ at $z = 0.6$.
This completeness threshold pushes to lower mass with increasing redshift,
dropping to $\sim4\times10^{14}\,\msun h^{-1}$ at $z=1$.
The size and redshift distribution of this catalog are in good agreement with expectations
based on our current understanding of galaxy clusters and cosmology.
In combination with other cosmological probes, we use this cluster catalog to improve
estimates of cosmological parameters.
Assuming a standard spatially flat \wCDM \  cosmological model,
the addition of our catalog to the WMAP 7-year results yields $\sigma_8 = 0.81 \pm 0.09$
and $w = -1.07 \pm 0.29$, a $\sim50\%$ improvement in precision on both parameters over WMAP7 alone.

\end{abstract}


\bigskip\bigskip

\section{Introduction}

\setcounter{footnote}{0}

The most massive dark matter halos to have formed so far
have characteristic masses of $10^{14}$ to $10^{15}$ solar masses.  
Although dark matter makes up the vast majority of the mass of these objects, 
most observation signatures result from baryons.  A small fraction 
of the baryons in these massive halos eventually cools to form stars and galaxies, and it 
was through the light from these galaxies that the most massive halos 
were first identified.  Because of this, we generally refer to these objects as galaxy
clusters, despite the small contribution of galaxies to their total mass.

Galaxy clusters are tracers of the highest peaks in the matter density field and, 
as such, their abundance is exponentially sensitive to the growth of structure over cosmic time.
A measurement of the abundance of galaxy clusters as a function of mass
and redshift has the power to constrain cosmological parameters to unprecedented 
levels \citep{wang98,haiman01,holder01b,battye03,molnar04,wang04,lima07}, assuming
that the selection criteria are well understood.
To usefully constrain the growth history of large scale structure, a sample of 
galaxy clusters must cover a wide redshift range. Furthermore, the observable 
property with which the clusters are selected should correlate strongly 
with halo mass, which is the fundamental quantity predicted from theory 
and simulations.
The thermal Sunyaev Zel'dovich \citep[SZ;][]{sunyaev72} signatures of galaxy clusters provide nearly these selection criteria.
Surveys of galaxy clusters based on the SZ effect have consequently been eagerly
anticipated for over a decade.  This paper presents the first cosmologically meaningful 
catalog of galaxy clusters selected via the thermal SZ effect.

\subsection{The Thermal SZ Effect}
The vast majority of known galaxy clusters have been
identified by their optical properties 
or from their X-ray emission.
Clusters of galaxies contain anywhere from several tens 
to many hundreds of galaxies, but these galaxies account for a small
fraction of the total baryonic mass in a cluster.  Most of the baryons in 
clusters are contained in the intra-cluster medium (ICM), the hot 
($\sim 10^7-10^8\,$K) X-ray-emitting plasma that pervades cluster environments.
\citet{sunyaev72} noted that this same plasma should also interact with
cosmic microwave background (CMB) photons via inverse Compton
scattering, causing a small spectral distortion of the CMB along the line
of sight to a cluster.  

The thermal SZ effect has been observed in 
dozens of known clusters (clusters previously identified in the optical 
or X-ray) over the last few decades \citep{birkinshaw99, carlstrom02}. 
However, it was not until very recently that the first previously unknown clusters 
were identified through their thermal SZ effect 
\citep{staniszewski09}.
This is mostly due to the small amplitude of the effect.
The magnitude of the temperature distortion at a given position on the
sky is proportional to the integrated electron
pressure along the line of sight. At the position of a massive galaxy
cluster, this fluctuation is only on the order of a part in $10^4$, or a
few hundred \uk \footnote{Throughout this work, the unit $\textrm{K}$
  refers to equivalent fluctuations in the CMB temperature, i.e.,~the
  level of temperature fluctuation of a 2.73$\,$K blackbody that would
  be required to produce the same power fluctuation. The conversion
  factor is given by the derivative of the blackbody spectrum,
  $\frac{dB}{dT}$, evaluated at 2.73$\,$K.}.
It is only with the current generation of large ($\sim$ kilopixel) detector arrays 
on 6-12~m telescopes \citep{fowler07, carlstrom09} that large areas of sky
are being surveyed to depths sufficient to detect signals of this amplitude.

A key feature of the SZ effect
is that the SZ surface brightness is insensitive to the redshift of the 
cluster.  As a spectral distortion of the CMB (rather
than an intrinsic emission feature), SZ signals redshift along
with the CMB. A given parcel of gas will
imprint the same spectral distortion on the CMB regardless of its
cosmological redshift, depending only on the electron density $n_e$
and temperature $T_e$. This makes SZ surveys an excellent technique for
discovering clusters over a wide redshift range.

Another aspect of the thermal SZ effect that makes it especially
attractive for cluster surveys is that the integrated thermal SZ flux
is a direct measure of the total thermal energy of the ICM.  The
SZ flux is thus expected to be a robust proxy for total cluster mass
\citep{barbosa96, holder01a, motl05}.

A mass-limited cluster survey 
across a wide redshift range provides a growth-based test of dark energy
to complement the distance based tests provided by supernovae
\citep{perlmutter99a, schmidt98}.
Recent results \citep[e.g.,][]{vikhlinin09, mantz10b}, have demonstrated the
power of such tests to constrain cosmological models and parameters.

\subsection{The SPT SZ Cluster Survey}
The South Pole Telescope (SPT) \citep{carlstrom09} is a 10-meter off-axis telescope optimized for
arcminute-resolution studies of the microwave sky. It is currently
conducting a survey of a large fraction of the southern sky with the
principal aim of detecting galaxy clusters via the SZ effect. In 2008, the
SPT surveyed $\sim 200\,$\sqdeg \ of the microwave sky with an array of 960
bolometers operating at 95, 150, and $220\,$GHz.
Using 40 \sqdeg of these data (and a small amount of overlapping data
from 2007), \citet{staniszewski09} (hereafter S09)
presented the first discovery of previously unknown clusters by their SZ
signature.
\citet{lueker10} (hereafter L10) used $\sim100\,$deg$^2$ \ of the 2008 survey
to measure the power spectrum of small scale temperature anisotropies in the CMB,
including the first significant detection of the contribution from the SZ secondary anisotropy.

In this paper we expand upon the results in S09 and present an
SZ-detection-significance-limited catalog of galaxy clusters
identified in the 2008 SPT survey.
Redshifts for 21 of
these objects have been obtained from follow-up optical imaging, the
details of which are discussed in a companion paper \citep{high10}.
Using simulated observations we characterize the SPT cluster selection
function --- the detectability of galaxy clusters in the survey as a function
of mass and redshift --- for the 2008 fields.
A simulation-based mass scaling relation allows us to compare the catalog
to theoretical predictions and place constraints on 
the normalization of the matter power spectrum
on small scales, $\sigma_8$,
and the dark energy equation of state parameter $w$.

This paper is organized as follows:
\S\ref{sec:obs-reduc} discusses
the observations, including data reduction, mapmaking, filtering,
cluster-finding, optical follow-up and cluster redshift estimation;
\S\ref{sec:results} presents the resulting cluster catalog;
\S\ref{sec:selection} provides a description of our estimate of the selection function;
\S\ref{sec:cosmology} investigates the sample in the context of our current
cosmological understanding and derives parameter constraints;
we discuss limitations and possible contaminants in
\S\ref{sec:systematics}, and we close with a discussion in
\S\ref{sec:discussion}.

For our fiducial cosmology we assume a spatially flat \LCDM \ model
(parameterized by $\Omega_b h^2$, $\Omega_c h^2$, $H_0$, $n_s$, $\tau$, and $A_{002}$)
with parameters consistent with the WMAP 5-year \LCDM \ best-fit results
\citep{dunkley09}\footnote{These parameters are sufficiently
  similar to the WMAP 7-year preferred cosmology \citep{larson10} that a
  re-analysis based on that newer work is not warranted.}, namely
$\Omega_M=0.264$, $\Omega_b=0.044$, $h=0.71$, $\sigma_8=0.80$. All references to cluster
mass refer to $M_{200}$, the mass enclosed within a spherical region
of mean overdensity $200\times\rho_{mean}$, where $\rho_{mean}$ is the
mean matter density on large scales at the redshift of the cluster.

\section{Observations, Data Reduction, Cluster Extraction, and Optical Follow-up}
\label{sec:obs-reduc}

\subsection{Observations}
\label{sec:obs}

The results presented in this work are based on observations 
performed by the SPT in 2008. 
\citet{carlstrom09} and S09 describe the details of
these observations; we briefly summarize them here.
Two fields were mapped to the nominal survey depth in 2008: one centered at right ascension (R.A.)
$5^\mathrm{h} 30^\mathrm{m}$, declination (decl.) $-55^\circ$ (J2000),
hereafter the \fiveh \ field; and one centered at R.A.  $23^\mathrm{h} 30^\mathrm{m}$,
decl.~$-55^\circ$, hereafter the \twentythreeh \ field.
Results in this paper are based on roughly 1500 hours of observing time
split between the two fields.  
The areas mapped with near uniform coverage were 
$91\,\sqdeg$ in the \fiveh \ field and $105\,\sqdeg$ in the \twentythreeh \ field.

This work considers only the $150\,$GHz data from the uniformly covered portions of the 2008 fields.
The detector noise for the $95\,$GHz detectors was very high for the 2008 
season,  and the $220\,$GHz observations were contaminated by the atmosphere at large 
scales where they would be useful for removing CMB fluctuations. Including 
these bands did not significantly improve the efficiency of cluster 
detections and they were not used in the analysis presented here.
The final depth of the \onefifty \ maps of the two fields is very similar,
with the white noise level in each map equal to $18\,\uk$-arcmin.

The two fields were observed using slightly different scan strategies. For the 
\fiveh \ field,
the telescope was swept in azimuth at a constant velocity ($\sim$0.25$^\circ$/s
on the sky at the field center) across the entire field then stepped in elevation,
with this pattern continuing until the whole field was covered.
The \twentythreeh \ field was observed using a similar strategy, except that 
the azimuth scans covered only one half of the field at any one time, switching
halves each time one was completed.
One consequence of this observing strategy was that a narrow strip in the middle
of the \twentythreeh \ field received twice as much detector time as the rest of the map.
The effect of this strip on our catalog is minimal and is discussed in \S\ref{sec:deepstrip}.

A single observation of either field lasted $\sim2\,$hours.  Between individual
observations, several short calibration measurements were performed, including
measurements of a chopped thermal source, $2$ degree elevation nods, and
scans across the galactic  HII regions RCW38 and MAT5a.  This series of regular
calibration measurements was used to identify detectors with good performance,
assess relative detector gains, monitor atmospheric opacity and beam
parameters, and model pointing variations.

\subsection{Data processing and mapmaking}
\label{sec:processing}
The data reduction pipeline applied to SPT data in this work is very similar to that used in S09.
Broadly, the pipeline for both fields consists of filtering the time-ordered data
from each individual detector, reconstructing the pointing for each detector,
and combining data from all detectors in a given observing band into a map
by simple inverse-variance-weighted binning and averaging. 

The small differences between the data reduction used in this work and that of
S09 are:
\begin{itemize}
\item
In S09, a 19th-order polynomial was fit and removed from each detector's 
timestream on each scan across the field. Samples in the timestream which mapped to
positions on the sky near bright point sources were excluded from the fit. 
A similar subtraction was performed here, except that a first-order polynomial was
removed, supplemented by
sines and cosines (Fourier modes). 
Frequencies for the Fourier modes were evenly spaced from $0.025\,$Hz to $0.25\,$Hz. 
This acts approximately as a high-pass filter in the R.A. direction with
a characteristic scale of $\sim 1 ^\circ$ on the sky.
\item
In S09, a mean across functioning detectors was calculated at each
snapshot in time and subtracted from each sample.
Here, both a mean and a
slope across the two-dimensional array were calculated at each
time and subtracted.
This acts as a roughly isotropic spatial high-pass filter, with a 
characteristic scale of $\sim 0.5^\circ$.
\item
As in \citet{vieira10}, a small pointing correction  ($\sim 5^{\prime\prime}$ on the sky) was applied, based on comparisons of radio source positions derived from SPT maps and positions of those sources in the AT20G catalog \citep{murphy10}.
\end{itemize}

The relative and absolute calibrations of detector response were performed as in L10.  
The relative gains of the detectors and their gain variations over time were estimated
using measurements of their response to a chopped thermal source.
These relative calibrations were then tied to an absolute scale through 
direct comparison of WMAP 5-year maps \citep{hinshaw09} to 
dedicated large-area SPT scans.
This calibration is discussed in detail in L10, and is
accurate to $3.6\%$ in the $150\,$GHz data.

\subsection{Cluster Extraction}
\label{sec:clusterfind}
The cluster extraction procedure used in this work for both fields is identical 
to the procedure used in S09, where more details can be found.

The SPT maps
were filtered to optimize detection of objects with morphologies 
similar to the SZ signatures expected from galaxy clusters, through the application of 
spatial matched filters 
\citep{haehnelt96,herranz02a,herranz02b,melin06}. In the spatial Fourier
domain, the map was multiplied by 
\begin{equation*}
\psi(k_x,k_y) = \frac{B(k_x,k_y) S(|\vec{k}|)}{B(k_x,k_y)^2 N_{astro}(|\vec{k}|) + N_{noise}(k_x,k_y)}
\end{equation*}
where $\psi$ is the matched filter, $B$ is the response of the
SPT instrument after timestream processing to signals on the sky,
$S$ is the assumed source template, and the noise power has been broken into
astrophysical ($N_{astro}$) and noise ($N_{noise}$) 
components.  For the source template, a projected spherical
$\beta$-model, with $\beta$ fixed to 1,
was used:
$$
\Delta T = \Delta T_0 (1+\theta^2/\theta_c^2)^{-1},
$$
where the normalization $\Delta T_0$ and the core radius $\theta_c$ are free parameters.

The noise power spectrum $N_{noise}$ includes contributions from atmospheric
and instrumental noise, while $N_{astro}$ includes power from primary and lensed
CMB fluctuations, an SZ background, and point sources.
The atmospheric and instrumental noise were estimated from jackknife maps as in S09,
the CMB power spectrum was updated to the lensed CMB spectrum from the WMAP5
best-fit \LCDM \ cosmology \citep{dunkley09}, the SZ background level was
assumed to be flat in $\ell(\ell+1)C_\ell$ with the amplitude taken from L10, and the
point source power was assumed to be flat in $C_\ell$ at the level given in \citet{hall10}.

To avoid spurious decrements from the wings of bright point sources, all
positive sources above a  given flux (roughly $7\,$mJy, or $5\,\sigma$ in a
version of the map filtered to optimize point-source signal-to-noise) were
masked to a radius of $4^\prime$ before the matched filter was applied.
Roughly 150 sources were masked in each field, of which $90$-$95 \%$
are radio sources.
The final sky areas considered after source masking were 82.4 and $95.1\,$\sqdeg \ for
the \fiveh \ and \twentythreeh \ fields respectively.

The maps were filtered for twelve different cluster scales, 
constructed using source templates with core radii $\theta_c$ evenly
spaced between $0.25^\prime$ to $3.0^\prime$.
Each filtered map $M_{ij}$, where $i$ refers to the filter scale and $j$ to the field,
was then divided into strips corresponding to distinct $90^\prime$
ranges in elevation. The noise was estimated independently in each
strip in order to account for the weak elevation dependence of the
survey depth. The noise in the $k^{th}$ strip of map $M_{ij}$,
$\sigma_{ijk}$, was estimated as the standard deviation of the map
within that strip.

Signal-to-noise maps $\widetilde{M}_{ij}$ were then constructed by dividing each strip $k$
in map $M_{ij}$ by $\sigma_{ijk}$.
SZ cluster decrements were identified in each map $\widetilde{M}_{ij}$ by a simple (negative)
peak detection algorithm similar to SExtractor \citep{bertin96}.
The highest signal-to-noise value associated with a decrement, across all filter scales,
was defined as \SN, and taken as the significance of a detection.
Candidate clusters were identified in the data down to \SN \ of 3.5, though this work
considers only the subset with \SN$\geq 5$.

These detection significances are robust against choice of source template:
the use of Nagai \citep{nagai07}, Arnaud \citep{arnaud10} or Gaussian templates in place of $\beta-$models
was found to be free of bias and to introduce negligible ($\sim 2\%$) scatter on recovered \SN.

\subsection{Optical Imaging and Spectroscopy}
\label{sec:optical}
Optical imaging was used for confirmation of candidates,
for photometric redshift estimation, and for cluster richness
characterization.  A detailed description of the coordinated optical
effort is presented in \citet{high10} and is summarized here.

The \fiveh \ and \twentythreeh \ fields were selected in part for
overlap with the Blanco Cosmology Survey \citep[BCS, see][]{ngeow09},
which consists of deep optical images using $g$, $r$, $i$ and $z$ filters.
These images, obtained from the Blanco 4m telescope at CTIO with
the MOSAIC-II imager in 2005--2007, were used when available. 
The co-added BCS images have 5$\,\sigma$ galaxy
detection thresholds of 24.75, 24.65, 24.35 and 23.05 magnitude
in $griz$, respectively.

For clusters that fell outside the BCS coverage region,
as well as for 5 that fell within, and for the unconfirmed candidate,
images were obtained using the twin $6.5\,$m Magellan
telescopes at Las Campanas, Chile.  
The imaging data on Magellan were obtained by taking successively 
deeper images until a detection of the early-type cluster galaxies was achieved,
complete to between $L^*$ and $0.4L^*$ in at least one band. 
The Magellan images were obtained
under a variety of conditions, and the Stellar Locus Regression
technique \citep{high09} was used to obtain precise colors and magnitudes. 

Spectroscopic data were obtained for a subset of the sample using the Low Dispersion Survey
Spectrograph on Magellan \citep[LDSS3, see][]{osip08} in longslit mode.
Typical exposures were $20$--$60$ minutes, with slit orientations
that contained the brightest cluster galaxy (BCG)
and as many additional red cluster members as possible.

Photometric redshifts were estimated using standard red sequence
techniques and verified using the spectroscopic subsample.
A red sequence model was derived from the work of \citet{bruzual03}, and
local overdensities of red galaxies were searched for near the cluster
candidate positions.
By comparing the resulting photo-$z$'s to spectroscopic redshifts within the subsample
of $10$ clusters for which spectroscopic data are available, H10 estimates
photo-$z$ uncertainty $\sigma_z$ as given in Table \ref{tab:catalog}, of order $3\%$ in $(1+z)$.

Completeness in red sequence cluster finding was estimated from mock catalogs.
At BCS depths, galaxy cluster completeness for the masses relevant for this sample
is nearly unity up to a redshift of one, above
which the completeness falls rapidly, reaching 50\% at about redshift
1.2, and 0\% at redshift 1.25.  At depths about a magnitude brighter
(corresponding to the depth of the Magellan observations of the
unconfirmed candidate, \S\ref{sec:cand_notes}), the
completeness deviates from unity at redshift $\sim 0.8$.

\section{Catalog}
\label{sec:results}

\begin{table*}
\begin{minipage}{\textwidth}
\centering
\caption{The SPT Cluster Catalog for the 2008 Observing Season} \small
\begin{tabular}{l cc cc cccc}
\hline\hline
\rule[-2mm]{0mm}{6mm}
Object Name			&
R.A.	& decl.		& \SN		& $\theta_c$	&
Photo-z		& $\sigma_z$	& Spec-z		& Opt	\\
\hline
SPT CL J0509-5342 \tablenotemark{\tabd}\tablenotemark{\tabdd}
										&77.336	&-53.705	&  6.61	& 0.50	& 0.47	& 0.04	& 0.4626	&BCS+Mag\\ 
SPT-CL J0511-5154	 \tablenotemark{a}			&77.920	&-51.904	&  5.63	& 0.50	& 0.74	& 0.05	& -		&Mag\\ 
SPT-CL J0516-5430	 \tablenotemark{\tabd}\tablenotemark{\tabdd}\tablenotemark{b}
										&79.148	&-54.506	&  9.42	& 0.75	& 0.25	& 0.03	& 0.2952	&BCS+Mag\\ 
SPT-CL J0521-5104 \tablenotemark{\tabdd}\tablenotemark{c}
										&80.298	&-51.081	&  5.45	& 1.00	& 0.72	& 0.05	& -		&BCS\\
SPT-CL J0528-5300  \tablenotemark{\tabd}\tablenotemark{d}
										&82.017	&-53.000	&  5.45	& 0.25	& 0.75	& 0.05	& 0.7648	&BCS+Mag\\ 
SPT-CL J0533-5005							&83.398	&-50.092	&  5.59	& 0.25	& 0.83	& 0.05	& 0.8810	&Mag\\ 
SPT-CL J0539-5744  \tablenotemark{\tabdd}		&85.000	&-57.743	&  5.12	& 0.25	& 0.77	& 0.05	& -		&Mag\\
SPT-CL J0546-5345	 \tablenotemark{\tabd}\tablenotemark{\tabdd}
										&86.654	&-53.761	&  7.69	& 0.50	& 1.16	& 0.06	& -		&BCS\\ 
SPT-CL J0551-5709	 \tablenotemark{\tabdd}\tablenotemark{e}
										&87.902	&-57.156	&  6.13	& 1.00	& 0.41	& 0.04	& 0.4230	&Mag\\ 
SPT-CL J0559-5249 \tablenotemark{\tabdd}		&89.925	&-52.826	&  9.28	& 1.00	& 0.66	& 0.04	& 0.6112	&Mag\\ 
SPT-CL J2259-5617	 \tablenotemark{\tabdd}\tablenotemark{f}
										&344.997	&-56.288	&  5.29	& 0.25	& 0.16	& 0.03	&0.1528	&Mag\\ 
SPT-CL J2300-5331	 \tablenotemark{\tabdd}\tablenotemark{g}
										&345.176	&-53.517	&  5.29	& 0.25	& 0.29	& 0.03	& -		&Mag\\ 
SPT-CL J2301-5546							&345.469	&-55.776	&  5.19	& 0.50	& 0.78	& 0.05	& -		&Mag\\ 
SPT-CL J2331-5051							&352.958	&-50.864	&  8.04	& 0.25	& 0.55	& 0.04	& 0.5707	&Mag\\ 
SPT-CL J2332-5358 \tablenotemark{\tabdd}\tablenotemark{h}
										&353.104	&-53.973	&  7.30	& 1.50	& 0.32	& 0.03	& -		&BCS+Mag\\ 
SPT-CL J2337-5942 \tablenotemark{\tabdd}		&354.354	&-59.705	& 14.94	& 0.25	& 0.77	& 0.05	& 0.7814	&Mag\\ 
SPT-CL J2341-5119							&355.299	&-51.333	&  9.65	& 0.75	& 1.03	& 0.05	& 0.9983	&Mag\\ 
SPT-CL J2342-5411							&355.690	&-54.189	&  6.18	& 0.50	& 1.08	& 0.06	& -		&BCS\\ 
SPT-CL J2343-5521							&355.757	&-55.364	&  5.74	& 2.50	& -		& -		& -		&BCS+Mag\\ 
SPT-CL J2355-5056							&358.955	&-50.937	&  5.89	& 0.75	& 0.35	& 0.04	& -		&Mag\\ 
SPT-CL J2359-5009							&359.921	&-50.160	&  6.35	& 1.25	& 0.76	& 0.05	& -		&Mag\\ 
SPT-CL J0000-5748							&0.250	&-57.807	&  5.48	& 0.50	& 0.74	& 0.05	& -		&Mag\\ 
\hline
\end{tabular}
\label{tab:catalog}
\begin{@twocolumnfalse}
\tablecomments{Recall that \SN \ is the maximum signal-to-noise obtained over the set of filter scales for each cluster. The $\theta_c$ (given in arcminutes) refer to the preferred filter scale, that at which \SN \ was found. Cluster positions in R.A. and decl. are given in degrees, and refer to the center of SZ brightness in map filtered at the preferred scale, calculated as the mean position of all pixels associated with the detection, weighted by their SZ brightness. The four rightmost columns refer to optical follow-up observations, giving the measured photometric redshift measurements and uncertainties of the optical counterpart, spectroscopic redshifts where available, and the source (BCS or Magellan) of the follow-up data.}
\tablenotetext{\tabd}{Clusters identified in S09. The cluster names have been updated (clusters were identified as SPT-CL 0509-5342, SPT-CL 0517-5430, SPT-CL 0528-5300, and SPT-CL 0547-5345 in S09) in response to an IAU naming convention request, an improved pointing model, and the updated data processing.}
\tablenotetext{\tabdd}{Clusters within $2^\prime$ of RASS sources \citep[RASS-FSC, RASS-BSC;][]{voges00, voges99}.}
\tablenotetext{a}{SCSO J051145-515430 \citep{menanteau10}.}
\tablenotetext{b}{Abell S0520 \citep{abell89}, RXCJ0516.6-5430 \citep{boehringer04}, SCSO J051637-543001 \citep{menanteau10}.}
\tablenotetext{c}{SCSO J052113-510418 \citep{menanteau10}.}
\tablenotetext{d}{SCSO J052803-525945 \citep{menanteau10}.}
\tablenotetext{e}{Abell S0552 \citep{abell89} is in the foreground, $5\arcmin$ away at z=0.09 (this redshift not previously measured).}
\tablenotetext{f}{Abell 3950 \citep{abell89}. Spectroscopic redshift from \citet{jones05b}.}
\tablenotetext{g}{Abell S1079 \citep[][redshift shown not previously measured]{abell89} .}
\tablenotetext{h}{SCSO J233227-535827 \citep{menanteau10}.}
\end{@twocolumnfalse}
\normalsize
\end{minipage}
\end{table*}

The resulting catalog of galaxy clusters, complete for \SN $\geq5$, is presented in Table \ref{tab:catalog}. 
Simulations (\S\ref{sec:sims}) suggest that this catalog should be highly complete above limiting mass and redshift thresholds, with relatively low contamination.

A total of 22 candidates were identified, for which optical follow-up 
confirmed and obtained redshift information on all but one.
Three clusters were previously known from X-ray and optical surveys,
three were previously reported from this survey by S09,
three were first identified in a recent analysis of BCS data by \citet{menanteau10}, 
and the remainder are new discoveries.
Detailed comparisons of the SPT and \citet{menanteau10} 
cluster catalogs and selection will be the subject of future work.

Thumbnail images of the signal-to-noise maps, $\widetilde{M}$, at the preferred
filter scale for each cluster are provided in Appendix \ref{app:thumbnails},
Figure \ref{fig:paper_thumbs}.
Signal-to-noise as a function of filter scale for each cluster is shown
in Appendix \ref{app:thumbnails}, Figure \ref{fig:paper_rcores}.

Estimates of cluster masses are possible with the aid of a scaling relation (below, \S\ref{sec:scaling_relation}), and are discussed in Appendix \ref{app:mass_estim}.

\subsection{Noteworthy Clusters}
\label{sec:cand_notes}
\paragraph{SPT-CL 2259-5617}
The SZ signal from this cluster is anomalously compact for such a low redshift object.
The cosmological analysis (\S\ref{sec:cosmology}) explicitly excludes all $z<0.3$
clusters, so this cluster not used in parameter estimation.

\paragraph{SPT-CL J2331-5051}
This cluster appears to be one of a pair of clusters at comparable redshift, likely undergoing a merger.
It will be discussed in detail in a future publication.
The fainter partner is not included in this catalog as its significance ($\SN=4.81$) falls below the detection threshold.

\paragraph{SPT-CL J2332-5358}
This cluster is coincident with a bright dusty point source which we identify in the 23h 
$220\,$GHz data.
Although the $150\,$GHz flux from this source could be removed with the aid
of the $220\,$GHz map, a multi-frequency analysis is outside the scope of the present work.
The impact of point sources on the resulting cluster catalog is discussed in \S\ref{sec:corr_ps}.

\paragraph{SPT-CL J2343-5521}
No optical counterpart was found for this candidate.
The field was imaged with both BCS and Magellan, and no cluster of galaxies was
found to a $5\,\sigma$ point source detection depth.
The simulated optical completeness suggests that this candidate is either
a false positive in the SPT catalog or a cluster at high redshift ($z \gtrsim 1.2$).
While the relatively high $\SN=5.74$ indicates a $\sim7\%$ chance
of a false detection in the SPT survey area (see discussion of
contamination below, \S\ref{sec:sf}),
the signal-to-noise of this detection exhibits peculiar behavior with $\theta_c$
(see Figure \ref{fig:paper_rcores}), preferring significantly larger scales than
any other candidate, consistent with a CMB decrement.
Further multi-wavelength follow-up observations are underway on
this candidate, and preliminary results indicate it is likely a false detection.

\subsection{Recovering integrated SZ flux}
\label{sec:sz_flux}
The optimal filter described in \S\ref{sec:clusterfind} provides an
estimate of the $\beta$-model normalization, $\Delta T_0$, and core size for
each cluster, based on the filter scale $\theta_c$ at which the
significance \SN \ was maximized. Assuming prior knowledge of the
ratio $\theta_{200}/\theta_c$ (where $\theta_{200}$ is the angle
subtending the physical radius $R_{200}$ at the redshift of the
cluster), one can integrate the $\beta$-profile to obtain an
estimate of the integrated SZ flux, $Y$. Basic physical
arguments and hydrodynamical simulations of clusters have demonstrated
$Y$ to be a tight (low intrinsic scatter) proxy for cluster mass
\citep{barbosa96, holder01a, motl05, nagai07, stanek10}.

In single-frequency SZ surveys, the primary CMB temperature
anisotropies provide a source of astrophysical contamination that
greatly inhibits an accurate measure of $\theta_c$
\citep{melin06}. The modes at which the primary CMB
dominates must be filtered out from the map, significantly
reducing the range of angular scales that can be used by the optimal filter
to constrain $\theta_c$. This range is already limited by the $\sim1^\prime$
instrument beam, which only resolves $\theta_c$ for the larger 
clusters.  Any integrated quantity will thus be
poorly measured. \citet{melin06} demonstrated that if the value of
$\theta_c$ can be provided by external observations, e.g., X-ray,
$Y$ can be accurately measured.

The inability to constrain $\theta_c$ can be seen in Figure
\ref{fig:paper_rcores}, where the highest signal-to-noise associated
with a peak is plotted against each filter scale. For several
clusters (for example, SPT-CL J0516-5430, SPT-CL J0551-5709, and
SPT-CL J2332-5358), the peak in
signal-to-noise associated with a cluster is very broad in 
$\theta_c$. Because of this confusion, we do not report $Y$ in this
work. Instead, as described below (\S\ref{sec:scaling_relation}), we
use detection significance as a proxy for mass.

Multi-frequency surveys are not in principle subject to
this limitation as the different frequencies can be combined to
eliminate sources of noise that are correlated between bands, thus
increasing the range of angular scales available for constraining
cluster profiles.

\section{SZ Selection Function}
\label{sec:selection}

In this section, we characterize the SPT cluster sample identified in
Table \ref{tab:catalog}.
Specifically, we describe the SPT cluster selection function in terms of 
the catalog completeness as a function of mass and redshift, and the 
contamination rate.
This selection function was
determined by applying the cluster detection algorithm described in
\S\ref{sec:clusterfind} to a large number of simulated SPT
observations. These simulations included the dominant astrophysical
components (primary and lensed CMB, cluster thermal SZ, and two families
of point sources), accounted for the effects of the SPT instrument
and data processing (the ``transfer function''), and contained
realistic atmospheric and detector noise.

\subsection{Simulated thermal SZ Cluster Maps}
\label{sec:sims}
Simulated SZ maps were generated using the method of \citet{shaw09},
where a detailed description of the procedure can be found. In brief,
the semi-analytic gas model of \cite{bode07} was applied to halos
identified in the output of a large dark matter lightcone
simulation. The cosmological parameters for this simulation were
chosen to be consistent with those measured from the WMAP 5-year data
combined with large-scale structure observations \citep{dunkley09},
namely $\Omega_M = 0.264$, $\Omega_b = 0.044$, and $\sigma_8 =
0.8$. The simulated volume was a periodic box of size 1 Gpc/h. The
matter distribution in 421 time slices was arranged into a lightcone
covering a single octant of the sky from $0 < z \leq 3$.

Dark matter halos were identified and gas distributions were
calculated for each halo using the
semi-analytic model of \citet{bode07}. This model assumes that
intra-cluster gas resides in hydrostatic equilibrium in the
gravitational potential of the host dark matter halo with a polytropic
equation of state. As discussed in \citet{bode07}, the most important
free parameter is the energy input into the cluster gas via
non-thermal feedback processes, such as supernovae and 
outflows from active galactic nuclei (AGN). 
This is set through the parameter $\epsilon_f$ such that
the feedback energy is $E_{f} = \epsilon_f M_* c^2$, where $M_*$ is
the total stellar mass in the cluster.
\citet{bode07} calibrate $\epsilon_f$ by comparing the model
against observed X-ray scaling relations for low redshift ($z<0.25$)
group and cluster mass objects. We note that the redshift range in
which the model has been calibrated and that encompassed by the
cluster sample presented here barely overlap;
comparison of the model to the SPT sample (as, for example, in \S\ref{sec:cosmology}) thus
provides a test of the predicted cluster and SZ signal evolution at high redshift.

For our fiducial model we adopt $\epsilon_f =
5\times10^{-5}$, however for comparison we also generate maps using
the `standard' and `star-formation only' versions of this model
described in \citet{bode09}. There are two principal differences between these
models. First, the stellar mass fraction $M_*/M_{\rm gas}$ is constant
with total cluster mass in the fiducial model, but mass-dependent in
the `standard' and  `star-formation' model.
Second, the amount of energy
feedback is significantly lower in the `standard' than model
than in the fiducial, and zero in the `star-formation' model.

From the output of each model, a 2-d image of SZ
intensity for each cluster with mass $M > 5 \times 10^{13}\,\msun
h^{-1}$ was produced by summing up the electron pressure along the
line of sight.  SZ cluster sky maps were constructed by projecting down
the lightcone, summing up the contribution of all the clusters along
the line of sight. Individual SZ sky maps were $10 \times 10$ degrees
in size, resulting in a total of 40 independent maps.  For each map,
the mass, redshift, and position of each cluster was recorded.

From SPT pointed observations of X-ray-selected clusters, \citet{plagge10} have demonstrated that cluster radial SZ profiles match the form of the ``universal'' electron pressure profile measured by \citet{arnaud10} from X-ray observations of massive, low-redshift clusters.
To complement
the set of maps generated using the semi-analytic gas model, SZ sky
maps were generated in which the projected form of the
\citet{arnaud10} pressure profile was used to generate the individual
cluster SZ signals.

\subsection{Point Source Model}
At $150\,$GHz and at the flux levels of interest to this analysis ($\sim 1$ to $\sim 10$~mJy), 
the extragalactic source population is expected to be primarily composed of two broad classes:  
sources dominated by thermal emission from dust heated by a burst of star formation, 
and sources dominated by synchrotron emission from AGN.  We refer to these two 
families as ``dusty sources'' and ``radio sources'' and include models for both in our
simulated observations.

For dusty sources, the source count model of 
\citet{negrello07} at $350\,$GHz was used. These counts are based on the physical
model of \citet{granato04} for high-redshift SCUBA-like sources and on a
more phenomenological approach for late-type galaxies (starburst plus normal
spirals). Source counts were estimated at $150\,$GHz by assuming
a scaling for the flux densities of $S_\nu \propto \nu^\alpha$, 
with $\alpha=3$ for high-redshift
protospheroidal galaxies and $\alpha=2$ for late-type galaxies. 
For radio sources, the \citet{dezotti05} model 
for counts at $150\,$GHz was used.
This model is consistent with the measurements of
\citet{vieira10} for the radio source population at $150\,$GHz.

Realizations of source populations were generated by sampling from
Poisson distributions for each population in bins with fluxes from
$0.01\,$mJy to $1000\,$mJy.  Sources were distributed in a random
way across the map.  Correlations between sources or with galaxy
clusters were not modeled, and we discuss this potential contamination
in \S\ref{sec:corr_ps}.

\subsection{CMB Realizations}

Simulated CMB anisotropies were produced by generating sky realizations based on
the power in the gravitationally lensed WMAP 5-year \LCDM \ CMB power spectrum.
Non-gaussianity in the lensed power was not modeled.

\subsection{Transfer Function}
\label{sec:transfunc}

The effects of the transfer function, i.e., the effects of the instrument 
beam and the data processing on sky signal, were emulated by producing 
synthetic SPT timestreams from simulated skies sampled
using the same scans employed in the observations.
The sky signal was convolved with the measured 
SPT $150\,$GHz beam, timestream samples were convolved with
detector time constants, and the SPT
data processing (\S\ref{sec:processing}) was performed on the 
simulated timestreams to produce maps.

Full emulation of the transfer function is a computationally intensive 
process; to make a large number of simulated observations, 
the transfer function was modeled as a 2D Fourier filter.
The accuracy of this approximation was measured by comparing recovered \SN \ of 
simulated clusters in skies passed through the full transfer function against
the \SN \ of the same clusters when the transfer function was approximated
as a Fourier filter applied to the map.
Systematic differences were found to be less than 1\% and on an 
object-by-object basis the two methods produced measured 
\SN \ that agreed to better than 3\%.

\subsection{Instrumental \& Atmospheric Noise}
\label{sec:noisesim}
Noise maps were created from SPT data by subtracting one half of each
observation from the other half. Within each observation, one
direction (azimuth either increasing or decreasing) was chosen at
random, and all data when the telescope was moving in that direction
were multiplied by -1.  The data were then processed and combined as
usual to produce a ``jackknife'' map which contained the full noise
properties of the final field map, but with all sky signal removed.

\subsection{The \twentythreeh \ Deep Strip}
\label{sec:deepstrip}
Due to the observing strategy employed on the \twentythreeh \ field, a
$\sim 1.5^\circ$ strip in the middle of that map contains significantly
lower atmospheric and instrumental noise than the rest of the map.  The
jackknife noise maps (\S\ref{sec:noisesim}) used in simulated observations naturally
include this deep strip, so any effects due to this feature are taken into
account in the simulation-based estimation of the average selection
function (\S\ref{sec:sf}) and scaling relation (\S\ref{sec:scaling_relation}) across the whole survey
region.  The cosmological analysis (\S\ref{sec:cosmology}) uses these averaged quantities;
simulated observations performed with and without a deep strip
demonstrated that any bias or additional scatter from using the averaged
quantities is negligible compared to the statistical errors.

\subsection{Completeness and Contamination}
\label{sec:sf}
Forty realizations of the 2008 SPT survey (two fields each) were simulated,
from which clusters were extracted and matched against input catalogs. 

\begin{figure}[]
\centering
\includegraphics[scale=0.58]{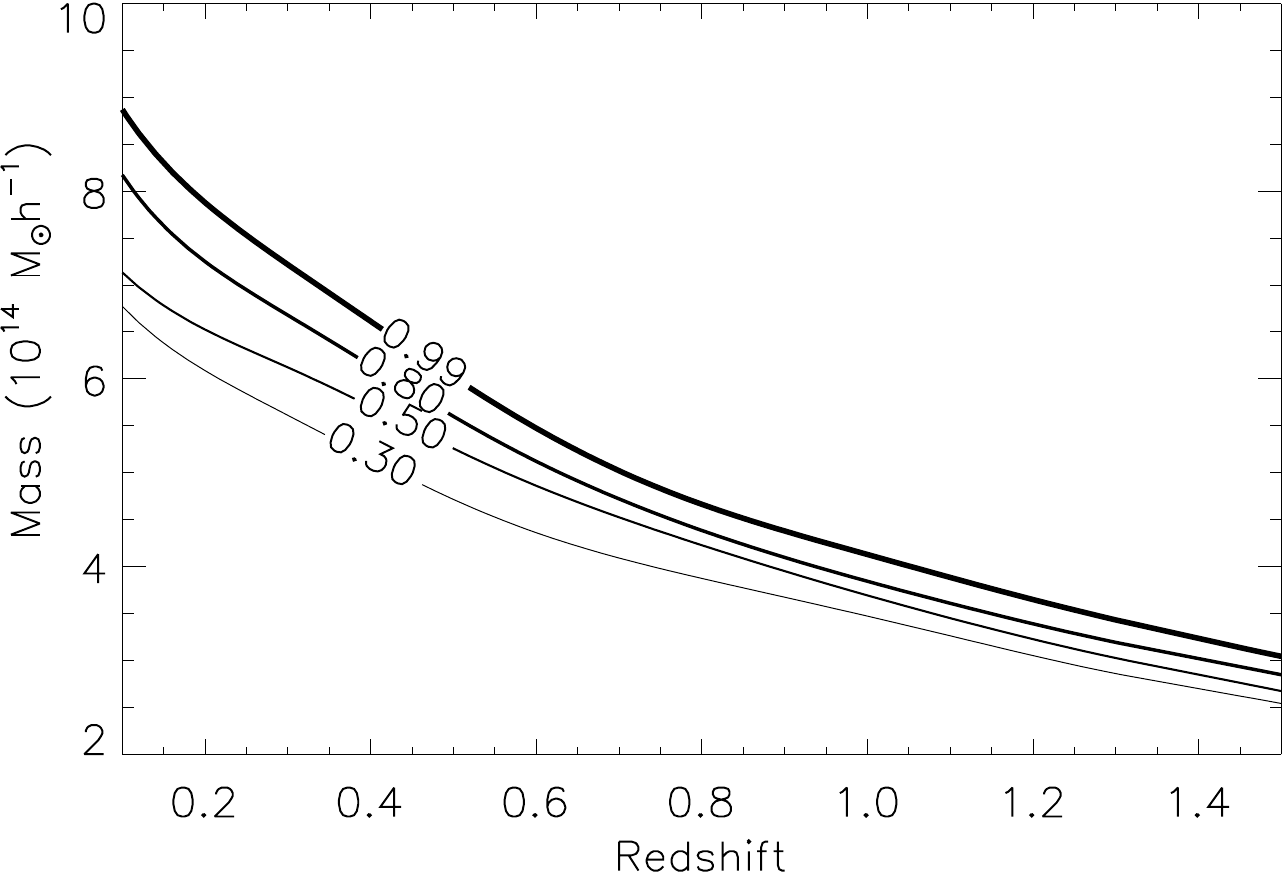}
  \caption[]{Simulated catalog completeness as a function of mass and redshift
  	for a significance cut of $\SN \geq 5$. The contours show lines of constant
	completeness.  From left to right, the lines represent 30, 50, 80 and 99\% completeness.
	The temperature and density of clusters at a given mass tends to increase with redshift,
	leading to the increased SZ flux and improved detectability of high-redshift
	clusters.
	The strong evolution below $z\sim0.3$ arises from reduced \SN \ on nearby
	clusters due to CMB confusion.
	Note that these contours are based on the fiducial simulations used in this work.
	Uncertainties in modeling (discussed in \S\ref{sec:scaling_relation}) can shift the
	position and shape of these contours coherently but significantly (of order $30\%$ in mass).
	\\
}
\label{fig:sz_completeness}
\end{figure}

Figure \ref{fig:sz_completeness} shows the completeness of the
simulated SPT sample, the fraction of clusters in simulated SPT maps
that were detected with $\SN \geq 5$, as a function of mass and redshift.
The exact shape and location of the curves in this figure depend on the
detailed modeling of intra-cluster physics, which remain uncertain.
The increase in SZ brightness (and cluster detectability) with increasing
redshift at fixed mass is due to the increased density and temperature of
high redshift clusters, and is in keeping with self-similar evolution.
At low redshifts ($z\lesssim0.3$), CMB confusion suppresses cluster detection
significances and drives a strong low-redshift evolution in the selection function.
These completeness curves were not used in the cosmological analysis (\S\ref{sec:cosmology}), where
uncertainties on the mass scaling relation (\S\ref{sec:scaling_relation}) account
for uncertainties in the modeling of intra-cluster physics.

\begin{figure}[]
\includegraphics[scale=0.88]{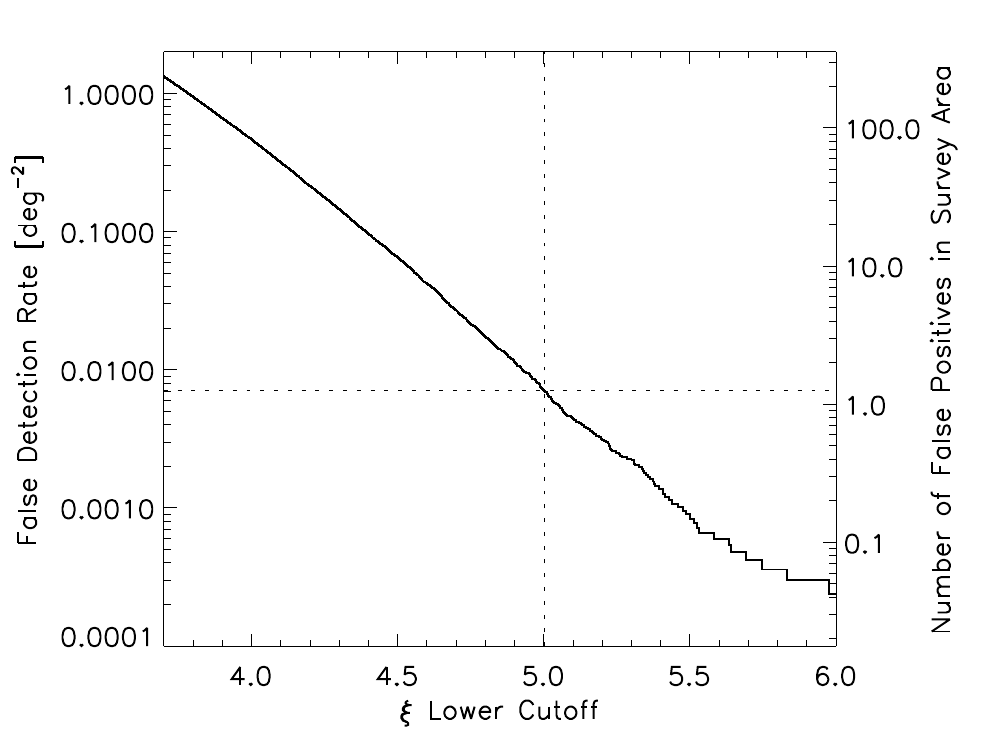}
  \caption[]{Simulated false detection rate, averaged across the survey area.
    The left axis shows the number density of false detections above a given
    \SN; the right axis shows the equivalent number of false
    detections within the combined \fiveh \ and \twentythreeh \ survey
    fields. The dotted lines show the $\SN\geq5$ threshold applied
    to the catalog, and the false detection rate at that threshold, $\sim1.2$
    across the full survey area.\\}
\label{fig:false_rate}
\end{figure}

The SZ sky was removed from simulations to estimate the rate of false positives
in the SPT sample.
Figure \ref{fig:false_rate} shows this contamination rate as a function of
lower \SN \ threshold, averaged across the survey area. A $\SN~\geq~5$ threshold 
leads to approximately one false detection within the survey area.

To test for biases introduced by an SZ background composed of low-mass systems,
a simulation was run including only SZ sources well below the SPT threshold, with
masses $M < 10^{14}\,\msun h^{-1}$. This background was found to have
negligible effect on the detection rate as compared to the SZ-free false detection simulation.

\subsection{Mass Scaling Relation}
\label{sec:scaling_relation}

As discussed in \S\ref{sec:sz_flux}, the integrated SZ flux $Y$ is poorly estimated
in this analysis and so is not used as a mass proxy.
However, the noise $\sigma_{ijk}$ measured in each elevation strip is relatively
even across the SPT maps, so it is possible to
work in the native space of the SPT selection function and use detection
significance \SN \ as proxy for mass.
Additional uncertainty and bias introduced by use of such a relation
(in place of, for example, a $Y$-based scaling relation) are small
compared to the Poisson noise of the
sample and the uncertainties in modeling intra-cluster physics.

The steepness of the cluster mass function in the presence of noise
will result in a number of detections that have boosted significance.
Explicitly, \SN \ is a biased estimator for
$\langle\SN\rangle$, the average detection significance of a given
cluster across many noise realizations.  An additional bias on \SN \ 
comes from the choice to maximize signal-to-noise across three free
parameters, R.A., decl. and $\theta_c$.
These biases make the relation between \SN \ and mass complex
and difficult to characterize.

In order to produce a mass scaling relation with a simple form, the unbiased
significance \nSN \ is introduced.
It is defined as the average detection signal-to-noise of a simulated
cluster, measured across many noise realizations, evaluated
at the preferred position and filter scale of that cluster
as determined by fitting the cluster in the absence of noise.

Relating \nSN \ and \SN \ is a two step process.
The expected relation between \nSN \ and $\langle\SN\rangle$ is 
derived and compared to simulated observations in Appendix 
\ref{app:nsn_deriv}, and found to be $\nSN = \sqrt{\langle\SN\rangle^2-3}$.
Given a known $\langle\SN\rangle$, the expected distribution in \SN \ 
is derived by convolution with a Gaussian of unit width.
The relation between \nSN \ and $\langle\SN\rangle$ is taken to be exact,
and was verified through simulations to introduce negligible additional
scatter; i.e., the scatter in the \nSN-\SN \ relation
is the same as the scatter in the $\langle\SN\rangle$-\SN \
relation, namely a Gaussian of unit width.

The scaling between $\nSN$ and $M$ is assumed to take the form of power-law
relations with both mass and redshift:
\begin{equation}
\label{eq:mass_scaling}
\nSN = A \left(\frac{M}{5\times10^{14}\,\msun h^{-1}}\right)^B \left(\frac{1+z}{1.6}\right)^C,
\end{equation}
parameterized by the normalization $A$, the slope $B$, and the redshift
evolution $C$. 
Appendix \ref{app:szscaling} presents a physical argument for the form of this relation,
along with the expected ranges in which the values of the parameters $B$ and $C$
are expected to reside based on self-similar scaling arguments.

Values for the parameters $A$, $B$, and $C$ were determined by fitting
Eq.~\ref{eq:mass_scaling} to a catalog of $\nSN > 1$ clusters
detected in simulated maps, using clusters with mass $M > 2\times
10^{14}\,\msun h^{-1}$ and in the redshift range $0.3 \leq z \leq
1.2$. This redshift range was chosen to match the SPT sample, while the
mass limit was chosen to be as low as possible without the sample being
significantly cut off by the $\nSN > 1$ threshold.
The best fit was defined as the combination of parameters that minimized
the intrinsic fractional scatter around the mean relation.

\begin{figure}[]
\includegraphics[scale=0.8]{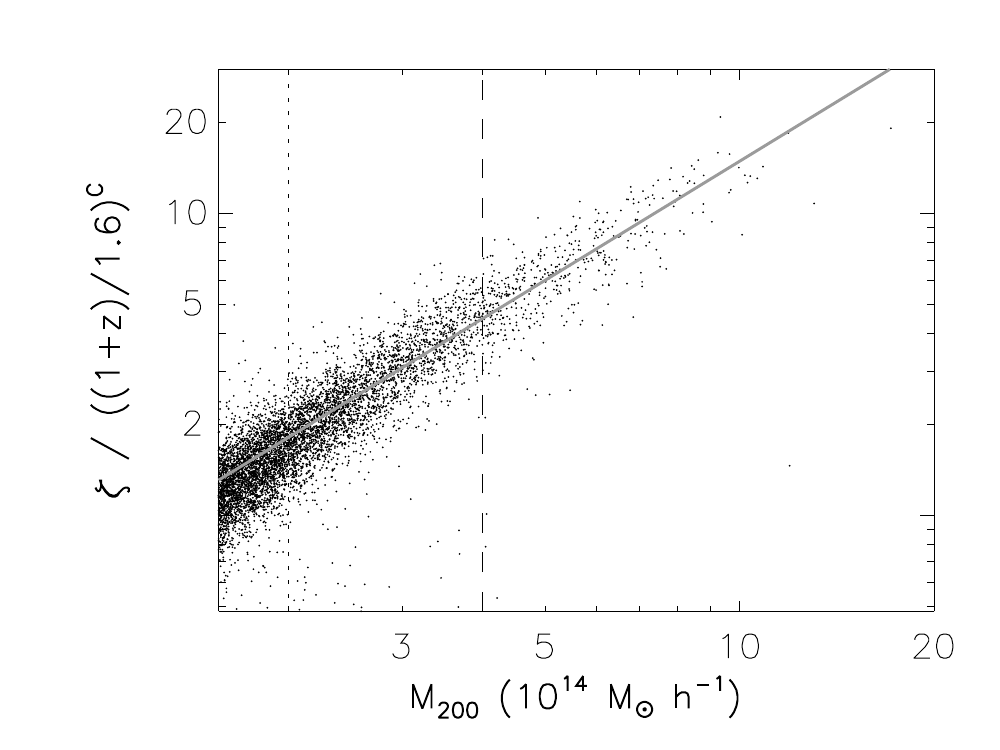}
  \caption[]{Mass-significance relation plotted over clusters identified in simulated maps.
  	The relation was fit to points with $M>2\times10^{14}\,\msun h^{-1}$, shown by the dotted line,
	and across a redshift range $0.3<z<1.2$.  Simulated clusters outside this redshift range are not
	included in this plot.
	The approximate lower mass threshold of the high-redshift end of the SPT sample ($M=4\times10^{14}\,\msun h^{-1}$)
	is shown by the dashed line.\\}
\label{fig:mass_scaling}
\end{figure}

Figure \ref{fig:mass_scaling} shows the best-fit scaling relation
obtained for our fiducial simulated SZ maps, where $A=6.01$, $B=1.31$,
and $C=1.6$. The intrinsic scatter was measured to be 21\% (0.21 in
$\ln(\nSN)$) and the relation was found to adhere to a power-law well
below the limiting mass threshold.
Over the three gas model realizations (\S\ref{sec:sims}), the best fit
value of $A$, $B$, and the intrinsic scatter were all found to vary by
less than 10\%, while the values of $C$ predicted by the `standard' and
`star-formation' models drop to $\sim1.2$.  For maps generated using the
electron pressure profile of \citet{arnaud10}, best-fit values of
$A=6.89$, $B=1.38$, $C = 0.6$ were found, with a 19\% intrinsic
scatter. The values of $A$ and $B$ remain within 15\% of the fiducial
model, although $C$ is significantly lower.
\citet{arnaud10} measured the pressure profile using a low-redshift
($z<0.2$) cluster sample and assume that the profile normalization
will evolve in a self-similar fashion.
The mass dependence of their pressure profile was determined using
cluster mass estimates derived from the equation of hydrostatic equilibrium;
simulations suggest that this method may underestimate the true mass
by $10-20\%$ \citep{rasia04, meneghetti10, lau09}.
We do not take this effect into account in our simulations -- doing so would reduce the value of $A$ by approximately 10\%.
The \citet{bode09} gas model is
calibrated against X-ray scaling relations measured from low-redshift
cluster samples \citep{vikhlinin06, sun09}, but assumes an evolving
stellar-mass fraction which may drive the stronger redshift evolution.

Based on these simulations, priors on the scaling relation parameters
($A$, $B$, $C$, scatter) were adopted, with conservative $1\,\sigma$ Gaussian
uncertainties of (30\%, 20\%, 50\%, 20\%) about mean values measured
from the fiducial simulation model.  These large uncertainties in
scaling relation parameters are the dominant source of uncertainty in
the cosmological analysis (\S\ref{sec:cosmology}) and mass estimation
(Appendix \ref{app:szscaling}). Furthermore, although the weakest
prior is on the redshift evolution, $C$, it is the uncertainty on the
amplitude $A$ that dominates the error budget on the measurement of
$\sigma_8$ (see \S\ref{sec:low_sz}).

It should be noted that at low redshift ($z\lesssim0.3$), such a
power-law scaling relation fails to fully capture the behavior
of the CMB-confused selection function.
The cosmological analysis below therefore excludes this region
during likelihood calculation.
The mass estimates presented in Appendix \ref{app:mass_estim}
may be biased low for low-redshift objects, although this effect
is expected to be small compared to existing systematic errors.

\section{Cosmological Analysis}
\label{sec:cosmology}
The 2008 SPT cluster catalog is an SZ-detection-significance-limited catalog. 
Simulated maps were used to calibrate the statistics of the relation between
cluster mass and detection significance, as well as the
impact of noise-bias and selection effects.
This relation was combined with theoretical mass functions to construct estimates of the 
number density of galaxy clusters as a function of the significance \SN \ and redshift,
to be compared to the SPT catalog.
Cosmological information from the SPT cluster catalog was combined
with information from existing data sets, providing improved
parameter constraints.

\subsection{Cosmological Likelihood Evaluation}
\label{sec:likelihood}
Evaluation of cosmological models in the context of the SPT catalog
requires a theoretical model that is capable of predicting the number 
density of dark matter halos as a function of both redshift and input cosmology. 
For a given set of cosmological parameters, the simulation-based
mass function of \citet{tinker08} was used in conjunction with
matter power spectra computed by CAMB \citep{lewis00}
to construct a grid of cluster number densities in the native $\SN$-z space of the SPT catalog:
\begin{itemize}
\item
A 2D grid of the number of clusters as a function of redshift and mass 
was constructed by multiplying the \citet{tinker08} mass function by the comoving
volume element.
The gridding was set to be very fine in both mass and redshift, with
$\Delta z=0.01$ and the mass binning set so that $\Delta \nSN=0.0025$ (see below).
The grids were constructed to extend beyond the
sensitivity range of SPT, $0.1 < z < 2.6$ and $1.8 < \nSN < 23$.
Extending the upper limits was found not to impact cosmological
results, as predicted number counts have dropped to negligible
levels above those thresholds.
\item
The parameterized scaling relation (\S\ref{sec:scaling_relation}) was used to 
convert the mass for each bin to unbiased significance \nSN \ for assumed
values of $A,B$ and $C$.
\item
This grid of number counts (in \nSN$-z$ space) was convolved with a
Gaussian in ln(\nSN) with width set by the assumed intrinsic scatter in the
scaling relation ($21\%$ in the fiducial relation).
\item
The unbiased significance $\nSN$ of each bin was
converted to an ensemble-averaged significance $\langle \SN \rangle$.
\item
This grid was convolved with a unit-width Gaussian in \SN \ to account for noise,
with the resulting grid in the native SPT catalog space, \SN$-z$.
\item
Each row (fixed \SN) of the \SN$-z$ grid which contained a cluster 
was convolved with a Gaussian with width set as
the redshift uncertainty for that cluster.
Photometric redshift uncertainties are given in Table \ref{tab:catalog}, 
and are described briefly in \S\ref{sec:optical} and in detail in \cite{high09};
spectroscopic redshifts were taken to be exact.
\item
A hard cut in \SN \ was applied, corresponding to the catalog 
selection threshold of $\SN \geq 5$. 
\item
An additional cut was applied, requiring $z\geq0.3$, to avoid
low-redshift regions where the power-law scaling relation fails to
capture the behavior of the CMB-confused selection function.
This cut excludes 3 low-redshift clusters from the cosmological
analysis, leaving 18 clusters plus the unconfirmed candidate,
whose treatment is described below.
\end{itemize}

The likelihood ratio for the SPT catalog was then constructed, 
as outlined in \citet{cash79}, using the Poisson probability,
$$
\mathcal{L}=\prod_{i=1}^N P_i=\prod_{i=1}^N \frac{e_i^{n_i}e^{-e_i}}{n_i !},
$$
where the product is across bins in \SN$-z$ space, $N$ is the number of bins,
 $P_i$ is the Poisson probability in bin $i$, and $e_i$ and $n_i$ are the 
fractional expected and integer observed number counts for that bin, respectively.

The unconfirmed candidate was accounted for by simultaneously
allowing it to either be at high redshift or a false detection.
Its contribution to the total likelihood was calculated as the union of 
the likelihoods for $n=0$ and $n=1$ within a large $z>1.0$ bin, the
redshift range corresponding to where the optical completeness for
this candidate's follow-up deviates from unity.

Ultimately, two sources of mass-observable scatter -- the intrinsic scatter
in the scaling relation, and the $1\,\sigma$ measurement noise --
were included in this analysis, along with redshift errors and systematic
uncertainties on scaling relation parameters.
Other sources of bias and noise (such as point source contamination,
\S\ref{sec:corr_ps}, and the mass function normalization described below)
are thought to be subdominant to these and were disregarded.

While \citet{tinker08} claim a very small ($<5\%$) uncertainty
in the mass function normalization, \citet{stanek10} have demonstrated that the
inclusion of non-gravitational baryon physics in cosmological simulations can
modify cluster masses in the range of the SPT sample by $\sim \pm10\%$ relative to
gravity-only hydrodynamical simulations.
The large $30\%$ uncertainty on the amplitude $A$ of the scaling relation
effectively subsumes such uncertainties in the mass function normalization.

This analysis does not account for the effects of sample variance
\citep{hu03a}; for the mass and redshift range of the SPT sample,
this is not expected to be a problem \citep{hu06}.
The SPT survey fields span of order 100 $h^{-1}$ Mpc, where the
galaxy cluster correlation function would be expected to be a few
percent or less \citep{bahcall03, estrada09}. This leads to clustering
corrections to the uncertainty on number counts on the order of
a few percent or less of the Poisson variance.

\subsection{Application to MCMC Chains}
\label{sec:mcmc}
The present SPT sample only meaningfully constrains a subset
of cosmological parameters, so to explore the cosmological
implications of the SPT catalog it is necessary to  include
information from other experiments.
Existing analyses of other cosmological data in the form of Markov Chain Monte Carlo (MCMC) chains
provide fully informative priors. 
These were importance sampled by weighting each set of 
cosmological parameters in the MCMC chain by the likelihood of the SPT 
cluster catalog given that set of parameters.

In this analysis, four MCMC chains were used to explore constraints on parameters:
the first two use only the 7-year data set from the WMAP experiment to explore
the standard spatially flat \LCDM \ and \wCDM \ cosmologies,
while the third explores \wCDM \ while adding data from
baryon acoustic oscillations (BAO) \citep{percival10} and supernovae (SNe) \citep{hicken09}.
These three chains were taken from the official WMAP
analysis\footnote{Chains available at http://lambda.gsfc.nasa.gov} \citep{komatsu10}.
The fourth chain was computed by L10 and allows for a direct comparison with that work.
It explores a spatially flat \LCDM \ parameter space based on the ``CMBall'' data
set: WMAP 5-year + QUaD \citep{brown09} + ACBAR \citep{reichardt09a} + SPT
(power spectrum measurements with $A_{SZ}$ as a free parameter; L10).

\begin{figure}[h]
\centering
\includegraphics[scale=0.45]{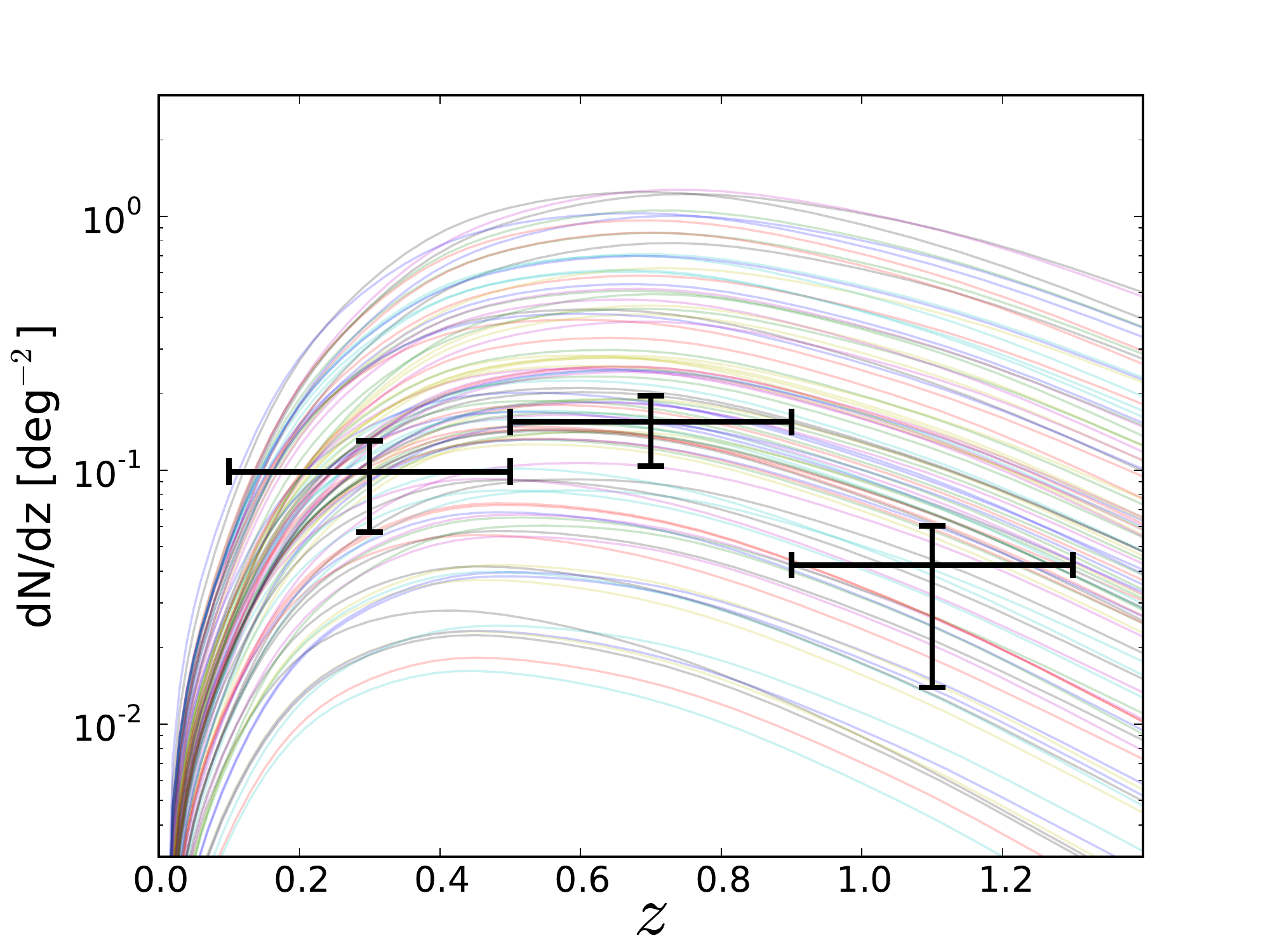}
  \caption[]{The SPT catalog, binned into 3 redshift bins (z=0.1-0.5, 0.5-0.9, 0.9-1.3), with number counts derived from 100 randomly selected points in the WMAP7 \wCDM \ MCMC chain overplotted. The SPT data are well covered by the chain and provide improved constraining power. The unconfirmed candidate is not included in this plot, and the binning is much coarser for display purposes than that used in the likelihood calculation (\S\ref{sec:likelihood}).\\}
\label{fig:sanity_plot}
\end{figure}

Figure \ref{fig:sanity_plot} shows the number density of clusters in the SPT
catalog, plotted over theoretical predictions calculated using the method described in
\S\ref{sec:likelihood}, for 100 random positions in the WMAP7-only MCMC chain.
The SPT data are adequately described by many cosmological models 
that are allowed by this data set, and the MCMC chains are 
well-sampled within the region of high probability.

Uncertainties in the scaling relation parameters were accounted for by
marginalizing over them:  at each step in the chain, the likelihood
was maximized across $A$, $B$, $C$ and scatter, subject to the priors
applied to each parameter, using a Newton-Raphson method.
The parameter values selected in this way at the highest likelihood point
in each MCMC chain are given in Table \ref{tab:marg_scaling_params}.
The fiducial values of
$B$ and the scatter appear consistent with those preferred by the
chains, while the preferred values of the normalization $A$ and
redshift evolution $C$ are both approximately 10\% lower than their
fiducial values.

This weaker-than-fiducial redshift evolution could come from a variety of
sources, and is consistent with other simulated models, e.g., the `standard'
and `star-formation' models, see \S\ref{sec:scaling_relation}.
Uncertainties in the redshift evolution are not a significant
source of error in this analysis:
recovered parameter values and uncertainties (\S\ref{sec:cosmo_results}) are found
to be insignificantly affected by widely varying priors on $C$.

\subsection{Cosmological Parameter Constraints}
\label{sec:cosmo_results}

\begin{figure*}[]\centering
\includegraphics[scale=0.75]{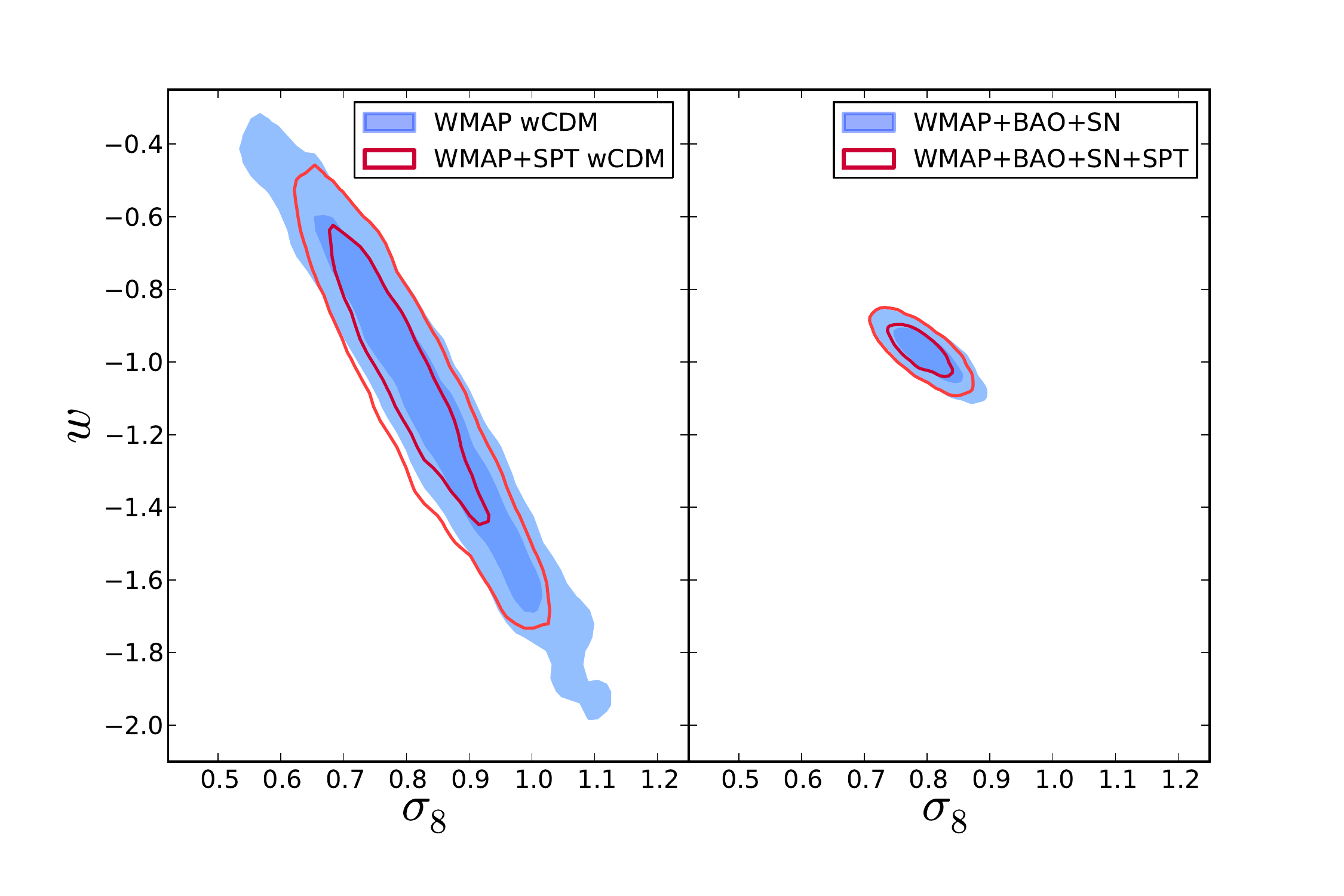}
  \caption[]{Likelihood contour plot of $w$ versus $\sigma_8$ showing $1\,\sigma$ and $2\,\sigma$ contours for several data sets. 
The left panel shows the constraints from WMAP7 alone (blue) and with the SPT cluster catalog included (red). The right panel shows show the full cosmological data set of WMAP7+SN+BAO (blue), and this plus the SPT catalog (red). The ability to constrain cosmological parameters is severely impacted by the uncertainties in the mass scaling relation, though some increase in precision is still evident.\\}
\label{fig:cont_s8w}
\end{figure*}

\begin{table*}[]
\begin{minipage}{\textwidth}
\centering
\caption{Cosmological Parameter Constraints} \small
\begin{tabular}{l cc}
\hline\hline
\rule[-2mm]{0mm}{6mm}
Chain				& $\sigma_8$		& $w$	\\
\hline

\LCDM \ WMAP7		& $0.801\pm0.030$	& $-1$					\\
\LCDM \ WMAP7+SPT 	& $0.791\pm0.027$	& $-1$					\\
\\		
\LCDM \ CMBall		& $0.794\pm0.029$	& $-1$					\\
\LCDM \ CMBall+SPT	& $0.788\pm0.026$	& $-1$					\\
\\
\wCDM \ WMAP7		& $0.832\pm0.134$	& $-1.118\pm0.394$			\\
\wCDM \ WMAP7+SPT	& $0.810\pm0.090$	& $-1.066\pm0.288$			\\
\\
\wCDM \ WMAP7+BAO+SNe		& $0.802\pm0.038$	& $-0.980\pm0.053$	\\
\wCDM \ WMAP7+BAO+SNe+SPT	& $0.790\pm0.034$	& $-0.968\pm0.049$	\\

\hline
\end{tabular}
\label{tab:cosmo_params}
\tablecomments{Mean values and symmetrized $1\,\sigma$ range for $\sigma_8$ and $w$,
as found from each of the four data sets considered, shown with and
without the weighting by likelihoods derived from the SPT cluster catalog.
The parameter best constrained by the SPT cluster catalog is $\sigma_8$.
CMB power spectrum measurements alone have a large degeneracy
between the dark energy equation of state, $w$, and $\sigma_8$.
Adding the SPT cluster catalog breaks this degeneracy and leads to an improved constraint on $w$.
The SPT catalog has negligible effect on other parameters in these chains
($\Omega_b h^2$, $\Omega_c h^2$, $H_0$,  $\tau$ and $n_s$).
}

\end{minipage}
\end{table*}

The resulting constraints on $\sigma_8$ and $w$ are given for all chains in Table \ref{tab:cosmo_params}.
The parameter best constrained by the SPT cluster catalog is $\sigma_8$.
CMB power spectrum measurements alone have a large degeneracy
between the dark energy equation of state, $w$, and $\sigma_8$.
Figure \ref{fig:cont_s8w} shows this degeneracy, along with the added constraints from the SPT cluster catalog.
Including the cluster results tightens the $\sigma_8$ contours and leads to an improved constraint on $w$.
This is a growth-based determination of the dark energy equation of state,
and is therefore complementary to dark energy measurements
based on distances, such as those based on SNe and BAO.

When combined with the \wCDM \ WMAP7 chain, the SPT data provide roughly
a factor of 1.5 improvement in the precision of $\sigma_8$ and $w$,
finding $0.81 \pm 0.09$ and $-1.07 \pm 0.29$, respectively.
Including data from BAO and SNe, these constraints tighten to
$\sigma_8 = 0.79 \pm 0.03$ and $w = -0.97 \pm 0.05$.

The dominant sources of uncertainty limiting these constraints
are the Poisson error due to the relatively modest size of the current
catalog and the uncertainty in the normalization $A$ of the mass scaling relation.
With weak-lensing- and X-ray-derived mass estimates of SPT clusters,
along with an order of magnitude larger sample expected from the full survey,
cosmological constraints from the SPT galaxy cluster survey will markedly improve.

\subsection{Amplitude of the SZ Effect}
\label{sec:low_sz}

\begin{figure}[]
\includegraphics[scale=0.45]{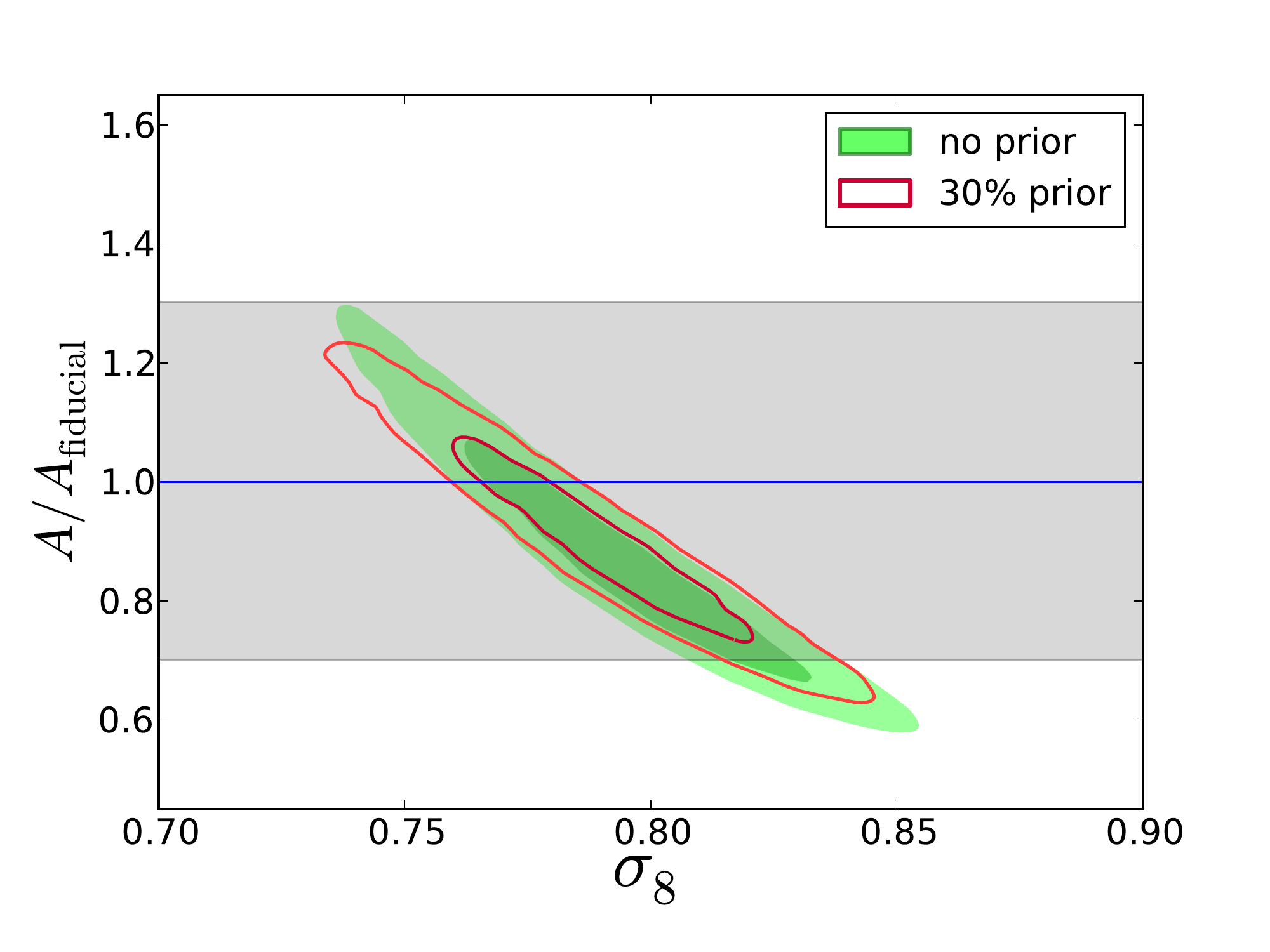}
  \caption[]{Degeneracy between $\sigma_8$ and SZ scaling relation amplitude $A$,
  	plotted without prior (green) and with a 30\% Gaussian prior (red) on $A$, for the \LCDM \
	WMAP5+CMBall MCMC chain. The Gaussian prior is shown ($\pm1\,\sigma$)
	by the gray band, with the fiducial relation amplitude shown by the blue line.
	This figure is analogous to Fig. 9 of L10, although that work dealt with SZ power,
	which is roughly proportional to the square of the amplitude being considered here.
	The prior is slightly higher than the preferred value; these results suggest that
	simulations may over-estimate the SZ flux in the high-mass, high-redshift
	systems contained in this catalog.\\}
\label{fig:s8A_prior}
\end{figure}

The value of the normalization parameter $A$ (which can be thought of as an ``SZ amplitude'') preferred
by the likelihood analysis was found to be lower than the fiducial
value, as shown in Figure \ref{fig:s8A_prior}.
The prior assumed on this parameter is sufficiently large that it
is not a highly significant shift; however, in light of 
the recent report by L10 of lower-than-expected SZ flux,
it is worth addressing.
The SPT cluster catalog results are complementary to the the results 
of the power spectrum analysis, in that 
the majority of the SZ power at the angular scales probed by L10
comes from clusters below the mass threshold of the cluster catalog.

Figure \ref{fig:s8A_prior} shows that the amplitude 
$A$ is strongly degenerate with $\sigma_8$.
The constraints provided by the SPT cluster catalog indicate either a value 
of $\sigma_8$ that is at the low end of the CMB-allowed distribution
(or equivalently an erroneously high mass function normalization),
or an over-prediction of SZ flux by the fiducial simulations.
If the fiducial amplitude is assumed, the best-fit $\sigma_8$ drops
from the WMAP5+CMBall value of $0.794\pm0.029$ to $0.775 \pm 0.015$. 
This value is anomalously low compared to recent results \citep[e.g.][]{vikhlinin09, mantz10b},
and in slight tension with the results of the power spectrum 
analysis of L10, where a still lower value of $\sigma_8=0.746 \pm 0.017$
was obtained for similar simulation models.\footnote{
	The fiducial thermal SZ simulation model used
	in this paper predicts a power spectrum that is in very
	close agreement with the fiducial model of L10, which
	was measured from the simulations of \citet{sehgal10}.}
The SZ amplitude parameter used in L10, $A_{sz}$, is roughly analogous to 
$A^2$ in the current notation.
When including the expected contribution from homogeneous reionization,
L10 found $A_{sz}=0.42 \pm 0.21$, in mild tension (at the $\sim 1 \sigma$ level) with the 
marginalized value of $(A/A_{fid})^2  = 0.79 \pm 0.30$ found in this analysis.

The fiducial simulations in this work use the semi-analytic gas model of
\citet{bode07,bode09}, which is calibrated against low-redshift
($z<0.25$) X-ray observations but has not previously been compared to
higher redshift systems.
One interpretation of these results is that
this model may over-predict the thermal electron pressure in
high-redshift ($z>0.3$) systems; this is not in conflict with the
low-redshift calibration of the model and suggests a weaker redshift
evolution in the SZ signal than predicted by the model.
Alternately, a combination of mass function normalization and
point source contamination could potentially account for the difference.

\section{Sources of Systematic Uncertainties}
\label{sec:systematics}

There are several systematic effects that might affect the utility of
the SPT cluster sample. For example, there remains large
uncertainty in the mapping between
detection significance and cluster mass. It is also possible that
strong correlations (or anti-correlations) between galaxy clusters and
mm-bright point sources are significant. We address these issues in this
section.

\subsection{Relation between SZ signal and Mass}

Theoretical arguments \citep{barbosa96, holder01a, motl05} suggest
that the SZ flux of galaxy clusters is relatively well understood.
However, there is very little high-precision empirical evidence to
confirm these arguments, and there are physical mechanisms that could
lead to suppressed SZ flux, such as non-thermal pressure support from
turbulence \citep{lau09} or non-equilibrium between protons and electrons
\citep{fox97, rudd09}.

Cluster SZ mass proxies (such as $Y$ and $y_0$, the integrated SZ flux and
amplitude of the SZ decrement, respectively)  depend linearly
on the gas fraction and the gas temperature.  There remain theoretical
and observational uncertainties in both of these quantities.
Estimates of gas fractions for individual clusters can disagree by
nearly 20\% \citep[e.g.,][]{allen08, vikhlinin06}, while theoretical and
observed estimates of the mass-temperature relation currently agree at
the level of 10-20\% \citep{nagai07}.  Adding these in quadrature
leads to uncertainties slightly below our assumed prior uncertainty of
30\%.

With the number counts as a function of mass, $dN/d\ln M$, scaling as
$M^{-2}$ or $M^{-3}$ for typical SPT clusters \citep{shaw10a}, a
$10\%$ offset in mass would lead to a 20-30\% shift in the number of
galaxy clusters.  With a catalog of 22 clusters, counting statistics
lead to an uncertainty of at least 20\%. Therefore, systematic offsets
in the mass scale of order 10\% will have a significant effect on
cosmological constraints, and the current $30\%$ prior
on $A$ will dominate Poisson errors.

A follow-up campaign using optical and X-ray observations
will buttress our current theory/simulation-driven
understanding of the SPT SZ-selected galaxy cluster catalog.

\subsection{Clusters Obscured by Point Sources}
\label{sec:corr_ps}

The sky density of bright point sources at $150\,$GHz is low enough --- on the order
of 1 $\mathrm{deg}^{-2}$ \citep{vieira10} --- that the probability of
a galaxy cluster being missed due to a chance superposition with a bright source  is negligible.
However, sources associated with clusters will preferentially fill in cluster SZ decrements.
Characterizing the contamination of cluster SZ measurements by member galaxies will be necessary 
to realize the full potential of the upcoming much larger SPT cluster catalog,
but the systematic uncertainty predicted here and in the literature 
is well below the statistical precision of the current sample; it is disregarded
in the current cosmological analysis (\S\ref{sec:cosmology}).

\subsubsection{Dusty Source Contamination}

Star formation is expected to be suppressed in cluster environments
\citep[e.g.,][]{hashimoto98}. \citet{bai07} measure the
abundance of infrared-luminous star-forming galaxies in a massive
($\gtrsim 10^{15} \msun$) cluster at $z=0.8$ to be far lower
relative to the field abundance than a simple mass scaling would predict:
the cluster volume that is hundreds of times overdense in mass is only 20 times overdense
in infrared luminosity.  A sphere at $z=1$ with a 1 Mpc radius and
infrared luminosity that is 20 times larger than the field would
produce $<0.1\,$mJy of emission at $150\,$GHz, according to the sub-mm
luminosity measurements of BLAST \citep{pascale09}(Pascale et al, 2009).  Even if the
IR overdensity evolves strongly with mass and redshift, we can expect
$\ll1\,$mJy  of contamination for the highest-redshift ($z \sim 1$)
clusters at the SPT mass threshold.  This corresponds to $\ll10\%$ of the
cluster SZ signal, which is far less than the uncertainty in the
normalization of cluster masses presented in this work.

Additionally, \citet[][in prep.]{keisler10} measures
the average $100$~$\mu \mathrm{m}$ flux of cluster
members from a sample of clusters at $\langle z\rangle=0.2$ and with masses
similar to those selected by SPT and, after extrapolating to $150\,$GHz and
allowing for strong redshift evolution in the infrared luminosity
function, constrains this contamination to be less than $10\%$ of the
cluster SZ signal. Again, this level of contamination is subdominant
to the uncertainty in the normalization of cluster masses presented in
this work.

\subsubsection{Gravitational Lensing}

Galaxy clusters can gravitationally lens sources
located behind them.  
Because gravitational lensing conserves surface brightness, this process 
cannot alter the mean flux due to the background sources when averaged 
over many clusters.
The background of sources is composed of both overdensities and
underdensities, leading to both positive and negative
fluctuations, relative to the mean, which will be gravitationally lensed.

We do not explicitly account for this effect in this work.
The unlensed fluctuating background of sources at $150\,$GHz is expected
to be small \citep{hall10} compared to both the experimental noise and intrinsic
scatter on the mass scaling relation, and lensing
only marginally increases the noise associated with these background
sources \citep{lima10}.
Within the context of the cosmological analysis, this additional noise term is 
expected to be small compared to the intrinsic scatter on the mass 
scaling relation. 

\subsubsection{Radio Source Contamination}
Galaxy clusters are known to host radio sources, but these correlated
sources are not expected to be a major contaminant at $150\,$GHz.
Calculations \citep{lin09} and explicit simulations \citep{sehgal10} demonstrate
that, even taking into account the expected correlation between clusters
and radio sources, these sources are not expected to significantly affect the SZ flux
in more than 1\% percent of galaxy clusters above $2 \times 10^{14} M_\odot$
at a redshift of $z \sim 0.5$ (where ``significantly" here means at the $\ge 20\%$ level). 

Simulations were also performed using knowledge of the radio source population
at $150\,$GHz from \citet{vieira10} and the cluster profiles that  maximize
the significance for the SPT clusters presented here.
Each profile between $\rcore = 0.25^\prime$ and $\rcore = 1.5^\prime$
(a range which encompasses all of the optically confirmed clusters in Table 
\ref{tab:catalog}) was scaled so that the filtered version of that 
profile would result in a $\SN=5$ detection in the 2008 SPT maps.
Point sources of a given flux were then added at a given radius from the profile center. 
These point-source-contaminated profiles were then convolved with the
transfer function, the matched filter was applied, and the resulting central
value was compared to the central value of the filter-convolved, uncontaminated
profile.
Clusters were found to suffer a systematic $\Delta\SN = 1$ reduction in
significance from a $2\,$mJy($5\,$mJy) source at $0.5^\prime$($1^\prime$) from the profile center.
This effect is nearly independent of core radius in the range of core 
radii probed.

The \citet{vieira10} radio source counts at $150\,$GHz indicate roughly $1.5$ per \sqdeg \ 
above $5\,$mJy, while the \citet{dezotti05} $150\,$GHz model predicts roughly
$3$ radio sources per \sqdeg \ above $2\,$mJy.\footnote{The \citet{vieira10} counts
	do not cover a low enough flux range to predict counts at $2\,$mJy,
	but the \citet{dezotti05} model is consistent with the \citet{vieira10}
	counts at all fluxes above $5\,$mJy, so counts from this model can
	confidently be extrapolated down a factor of $2.5$ in flux.}
If there were no correlation between clusters and radio sources, the clusters contained in
the SPT catalog should have a $0.14\%$  $(0.03\%)$ chance of
incurring an error of $\Delta\SN\geq1$ from a $\ge 5\,$mJy ($2-5\,$mJy) source.

Furthermore, using $30\,$GHz observations of a sample of clusters ranging from $0.14 < z < 1.0$,
\citet{coble07} find the probability of finding a radio source near a cluster to be
$8.9^{+4.3}_{-2.8}$ ($3.3^{+4.1}_{-1.8}$) times the background rate 
when using a $0.5^\prime$($5^\prime$) radius.
From these results, it can be estimated that roughly $1\%$ of SPT-detected clusters would suffer
an error of $\Delta\SN\geq1$ from radio source contamination.  This is in very close agreement
with the predictions from \citet{lin09} and \citet{sehgal10}.

\section{Discussion}
\label{sec:discussion}
We have presented the first cosmologically significant SZ-selected galaxy
cluster catalog, characterized the selection function, and performed a
preliminary cosmological analysis to both demonstrate the general 
consistency of the catalog with current understanding of cosmology
and provide improved constraints on cosmological parameters.
This is an important step toward exploiting the potential of SZ-selected
galaxy clusters as a powerful cosmological tool.

Using single-frequency data taken in 2008 with SPT, a total of 22 candidates
were identified, all but one of which were optically confirmed as galaxy clusters \citep{high10}.
Of these 21 clusters, three were previously known from optical and/or X-ray surveys, 
three were new SPT detections reported in S09,
three were first identified from BCS data by \citet{menanteau10},
and 12 are new discoveries.

Simulations were used to calibrate the selection function of the survey and measure a
scaling between SPT detection significance and mass.
These simulations indicate that SZ detection
significance traces mass with little ($\sim20\%$) intrinsic scatter,
making SZ surveys well suited to selecting mass-limited catalogs of galaxy clusters.

As a demonstration of the constraining power of the survey,
the SPT cluster catalog was used to refine estimates of cosmological parameters, including
the dark energy equation of state, $w$, and the normalization of the matter power spectrum
on small scales, $\sigma_8$. 
Using \wCDM \ MCMC chains derived from the WMAP 7-year data combined with the
SPT cluster catalog,  the best-fit values were $w=-1.07 \pm 0.29$ and
$\sigma_8=0.81 \pm 0.09$, a factor of roughly 1.5 improvement in precision compared to the WMAP7 constraints alone. 
When combined with other cosmological data sets (baryon acoustic oscillations and supernovae),
the SPT cluster catalog improves precision on these parameters by $\sim 10\%$.

These results can be compared to those of \citet{vikhlinin09} and \citet{mantz10b}, 
who performed a similar analysis using large samples of clusters drawn from X-ray surveys.
The SPT results are less precise: in combination with 
various cosmological data sets, \citet{vikhlinin09} find nearly 4 times 
tighter constraints on $\sigma_8$, while \citet{mantz10b} are more precise
by nearly a factor of 2.  This is not surprising: both X-ray analyses had
significantly larger cluster samples and smaller stated 
uncertainty in the mass scaling relation.
The weaker parameter constraints found from the SPT cluster catalog
are a direct result of uncertainties in the mass scaling relation,
which derive from uncertainties modeling intra-cluster physics.

The fiducial thermal SZ simulation model assumed here was shown to produce some
tension between the analysis presented here and contemporary cosmological results.
This may be explained by a variety of factors.
The value of $\sigma_8$ may be lower than currently favored, or equivalently the
normalization of the \citet{tinker08} mass function may be erroneously high.
Alternatively, current simulations, while reproducing observed low-redshift X-ray
observations, may over-estimate SZ flux in higher redshift systems, implying
missing physics in the semi-analytic gas modeling \citep[for example, a
non-negligible amount of non-thermal pressure support at higher redshifts, ][]{shaw10b}.
The observed SZ signal could potentially be contaminated by an
increasing incidence of point-source emission at high redshift
although the arguments presented in \S\ref{sec:corr_ps} suggest
point-source contamination is unlikely to be wholly responsible.
L10 also found lower-than-anticipated power from the SPT measurement
of the SZ power spectrum, consistent with many of these scenarios.
A concerted, multi-wavelength program  aimed at studying high redshift
clusters should help to resolve these issues.

The SPT catalog presented here is based on less than 1/3 of the current data
and roughly 1/10 of the full SPT survey. The large multifrequency  
SPT survey, combined with X-ray and/or weak lensing mass estimates of 
a subsample of SZ-selected galaxy clusters, should allow an order of magnitude
improvement in the precision of $\sigma_8$ and $w$ measurements.

\acknowledgments

The SPT team gratefully acknowledges the contributions to the design
and construction of the telescope by S.\ Busetti, E.\ Chauvin,
T.\ Hughes, P.\ Huntley, and E.\ Nichols and his team of iron
workers. We also thank the National Science Foundation (NSF) Office of
Polar Programs, the United States Antarctic Program and the Raytheon
Polar Services Company for their support of the project.  We are
grateful for professional support from the staff of the South Pole
station. We thank H.-M.\ Cho, T.\ Lanting, J.\ Leong, W.\ Lu, M.\ Runyan, D.\ Schwan,
M.\ Sharp, and C.\ Greer for their early contributions to the SPT
project and J.\ Joseph and C.\ Vu for their contributions to the
electronics.
We acknowledge S.\ Alam, W.\ Barkhouse, S.\ Bhattacharya, L.\ Buckley-Greer, S.\ Hansen, H.\ Lin, Y-T Lin, C.\ Smith and D.\ Tucker for their contribution to BCS data acquisition, and we acknowledge the DESDM team, which has developed the tools we used to process and calibrate the BCS data

We acknowledge the use of the Legacy Archive for Microwave 
Background Data Analysis (LAMBDA). Support for LAMBDA is provided 
by the NASA Office of Space Science.
This research was facilitated in part by allocations of time on the COSMOS
supercomputer at DAMTP in Cambridge, a UK-CCC facility supported by
HEFCE and PPARC.
This work is based in part on observations obtained at the Cerro
Tololo Inter-American Observatory, and the Las Campanas Observatory. 
CTIO is operated by the Association of Universities for Research in Astronomy (AURA),
Inc., under cooperative agreement with the National Science Foundation (NSF). 

The South Pole Telescope is supported by the National Science
Foundation through grants ANT-0638937 and ANT-0130612.  Partial
support is also provided by the NSF Physics Frontier Center grant
PHY-0114422 to the Kavli Institute of Cosmological Physics at the
University of Chicago, the Kavli Foundation and the Gordon and Betty
Moore Foundation. This research used resources of the National Energy
Research Scientific Computing Center, which is supported by the Office
of Science of the U.S. Department of Energy under Contract
No. DE-AC02-05CH11231.
This work is supported in part by the Director, Office of Science,  
Office of High Energy Physics, of the U.S. Department of Energy under  
Contract No. DE-AC02-05CH11231.
The McGill group acknowledges funding from
the National Sciences and Engineering Research Council of Canada, the
Quebec Fonds de recherche sur la nature et les technologies, and the
Canadian Institute for Advanced Research.
Partial support was provided by NSF grant MRI-0723073.
The following individuals acknowledge additional support:
A. Loehr and B. Stalder from the Brinson Foundation,
B. Benson and K. Schaffer from KICP Fellowships,
J. McMahon from a Fermi Fellowship, 
R. Foley from a Clay Fellowship,
D. Marrone from Hubble Fellowship grant HF-51259.01-A,
M. Brodwin from the Keck Foundation,
Z. Staniszewski from a GAAN Fellowship, and A.T. Lee from
the Miller Institute for Basic Research in Science, University of California Berkeley.

Facilities: Blanco (MOSAIC II), Magellan:Baade (IMACS), Magellan:Clay (LDSS2)

\bibliography{../../BIBTEX/spt.bib}

\newpage
\appendix

\section{Thumbnail images of cluster SZ decrements}
\label{app:thumbnails} 

\begin{figure}[h]
\centering
\includegraphics[scale=.9]{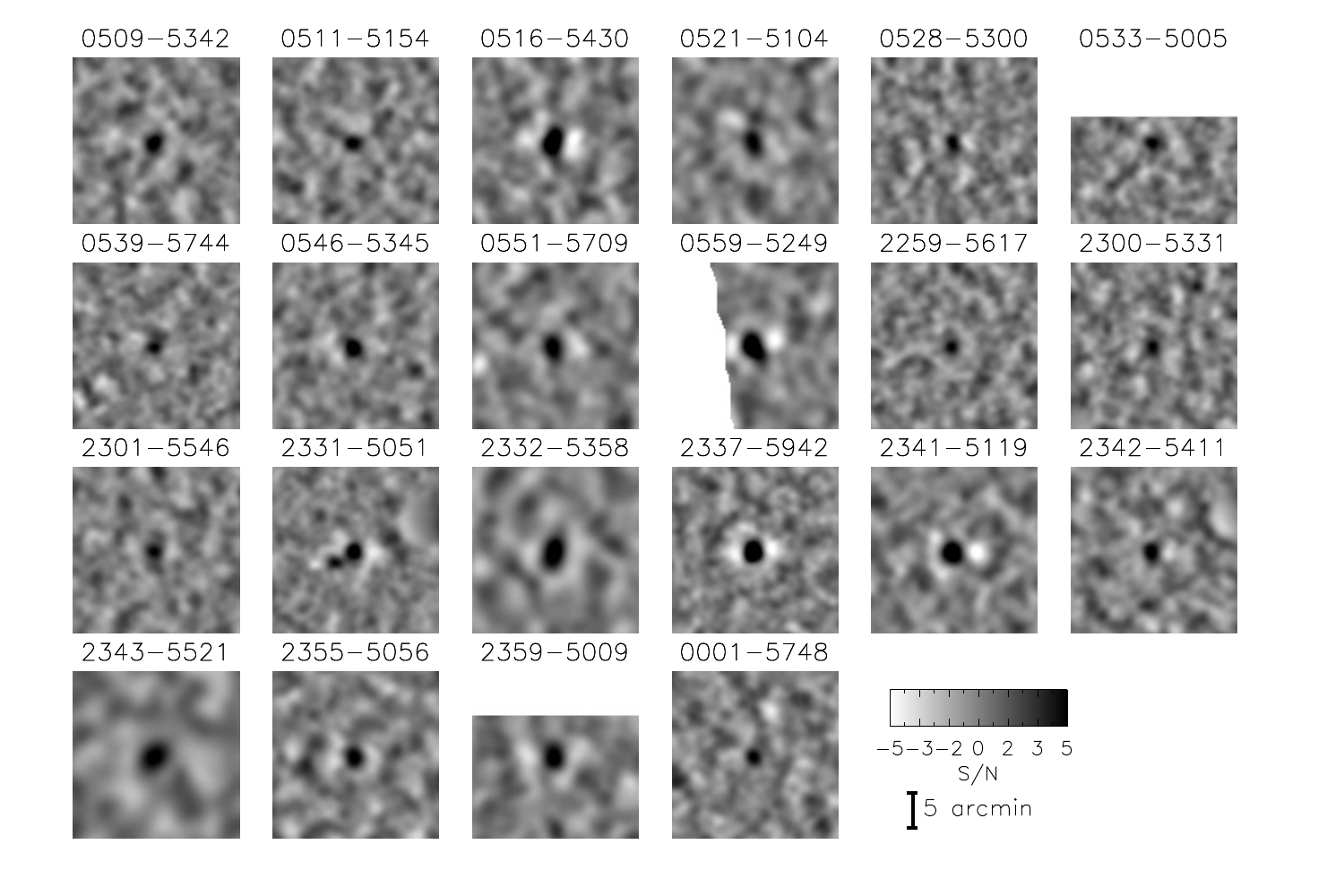}
  \caption[]{Thumbnail cutouts of filtered maps, 25' x 25', 
  	of the highest significance filter scale for each cluster,
	centered on the cluster position.
	The color scale and spatial extent of all maps here are the same,
	and the irregular shape of some images shows the edge of the masked (even coverage) map.
	The large blur in the upper right of SPT-CL J2331-5051 is a point source mask.
	}
\label{fig:paper_thumbs}
\end{figure}

\begin{figure}[h]
\centering
\includegraphics[scale=.9]{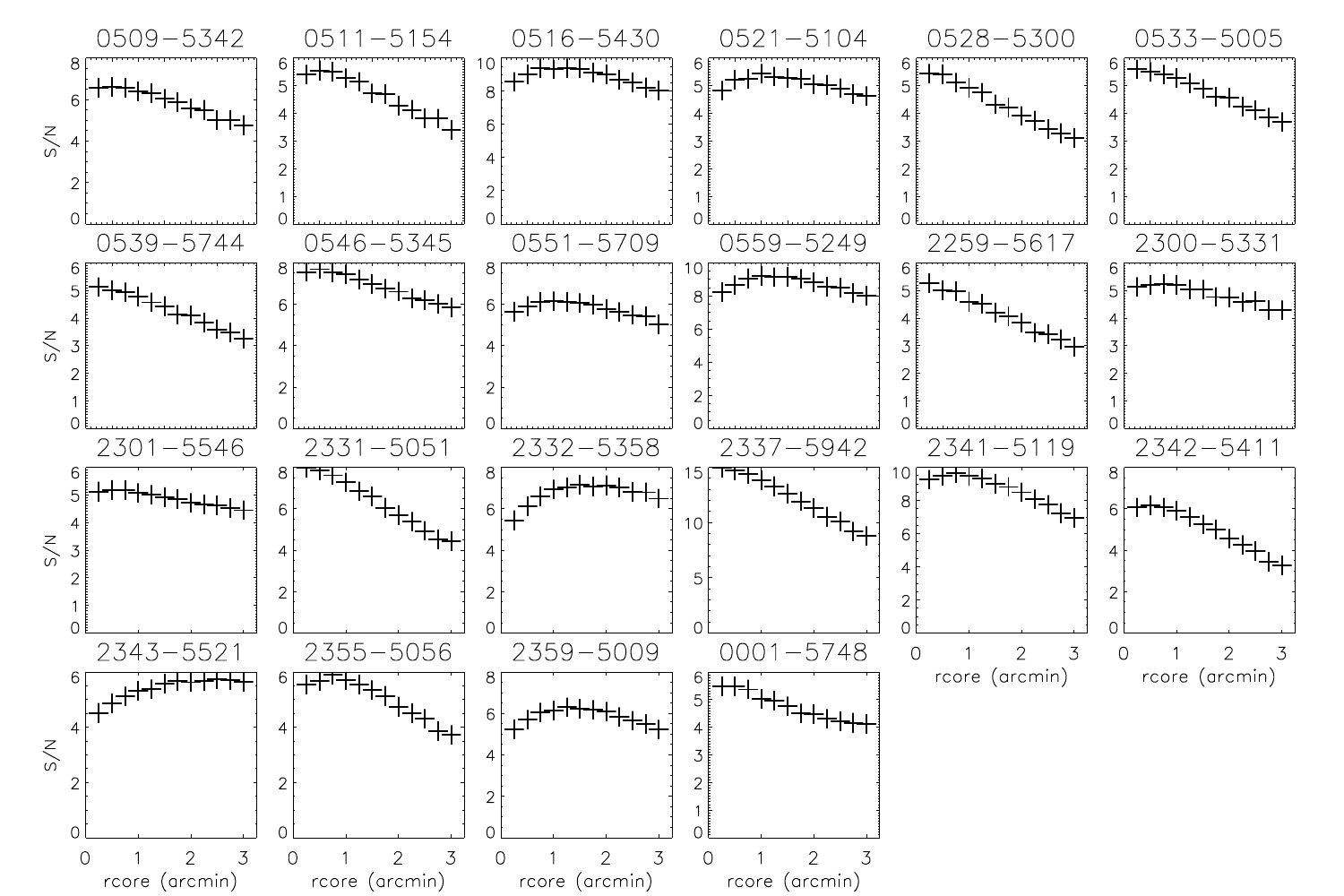}
  \caption[]{The measured significance across filter scales for all clusters.
  	The maximum value of each curve gives \SN \ for that candidate.
  	The significance axes have been allowed to float for each panel.
	Adjacent points are highly correlated, and as discussed in \S\ref{sec:sz_flux},
	the best-fit $\theta_c$ is poorly constrained.
	}
\label{fig:paper_rcores}
\end{figure}

\newpage

\section{Unbiased Significance \nSN}
\label{app:nsn_deriv}
The freedom to maximize the SPT significance \SN \ across three parameters (R.A., decl., and $\theta_c$) in the presence of a noise field will tend to raise the amplitude of the observed peak. That is, the ensemble average of \SN \ across many noise realizations, $\langle\SN\rangle$, will be boosted by some amount as compared to the unbiased significance \nSN, which is measured without these degrees freedom.

We can consider the SPT significance $\xi$ as analogous to 
a $\chi^2$ of the best-fit model relative to zero signal, with the signal
to noise being comparable to $\sqrt{\chi^2}$. By allowing
three degrees of freedom in a usual $\chi^2$ fit, we expect that the minimum
$\chi^2$ will
typically be smaller than the corresponding one for a fit with no
degrees of freedom, i.e., at the true location using the true filter size.
The typical difference is simply the number of fit parameters, in our
case three: $\langle\SN\rangle^2-\nSN^2=3$.

\begin{figure}[h]
\centering
\includegraphics[scale=1.]{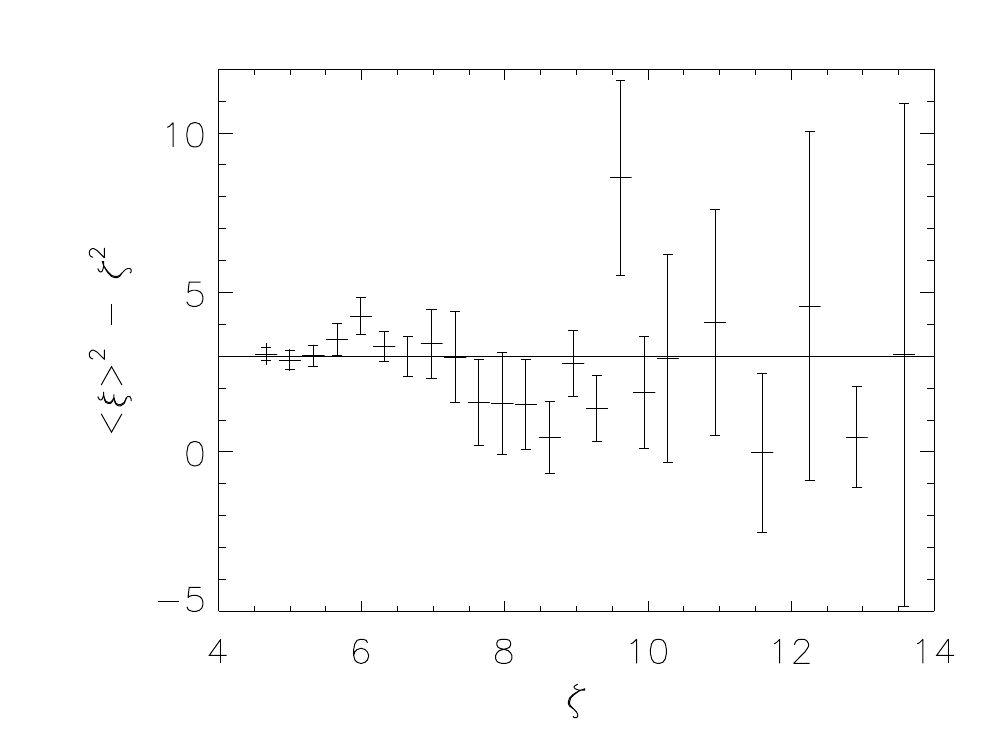}
  \caption[]{The quadratic difference between measured \nSN \ and $\langle\SN\rangle$ (measured across 25 realizations of all noise terms), binned and plotted against a wide range of \nSN. These data are consistent with $\langle\SN\rangle^2-\nSN^2=3$, with $\chi^2 = 27$ for 23 degrees of freedom.}
\label{fig:zeta_xi}
\end{figure}

We test the relation on simulated data, adding 25 different realizations of each noise term (CMB, point sources, instrumental and atmospheric) to the same SZ sky, and explicitly measuring $\langle\SN\rangle$ for many simulated clusters in these maps. We next simulate and measure \nSN \ for the same patch of sky and compare the two measures of significance. The results of these simulations are shown in Figure \ref{fig:zeta_xi}, and are consistent with the hypothesized relation
 with $\chi^2 = 27$ for 23 degrees of freedom.

We adopt this relation for the cosmological analysis and mass estimation,
$$
\nSN \approx \sqrt{\langle\SN\rangle^2 - 3},
$$
and find that any residual behavior not modeled by this relation is small enough to be well sub-dominant to uncertainties on the scaling relation amplitude $A$ and Poisson noise within the catalog. 
Within the SPT catalog this is a small effect:
for $\langle\SN\rangle = 5$(10) objects, the difference between \nSN \ and $\langle\SN\rangle$ is 6\%(1\%).

\newpage

\section{Mass-Significance Scaling Parameters}
\label{app:szscaling}

We have assumed that the scaling between \nSN \ and $M$ takes the form
\begin{equation}
\nSN = A \left(\frac{M}{5\times10^{14}\,\msun h^{-1}}\right)^B
\left(\frac{1+z}{1.6}\right)^C \;\;,
\label{eqn:mass_scaling_app}
\end{equation}
where cluster mass $M$ is defined in terms of a spherical overdensity
relative to the mean density 
at a given redshift $z$ and the pivot points in this relation were 
chosen to roughly match the mean mass and redshift of clusters in the 
SPT sample. Estimates for $B$ and $C$ were obtained from extensive
simulations; in this appendix we explore various limiting cases to
better understand what values are expected. For the cosmological
analysis we use priors informed by our simulations. 

The amplitude of the SZ effect through the center of a cluster (the
central decrement) is expected to scale as \citep{holder99a}
\begin{equation}
y_0 \propto M (1+z)^3 \;\;,
\end{equation}
so if \nSN \ is proportional to $y_0$ we would expect $B = 1$ and
$C = 3$. 

The integrated SZ flux $Y$ $(= \int y d\Omega)$ scales with mass as
\citep{barbosa96} 
\begin{equation}
Y \propto M^{5/3} (1+z) / D_A(z)^2 \;\;,
\end{equation}
where $\Omega$ is the solid angle subtended by a cluster with radius
$R$ at redshift $z$ and $D_A(z)$ is the angular diameter distance to
redshift $z$.

For unresolved detections, where the signal
to noise is simply proportional to $Y$ divided by the noise per resolution
element,  we would expect
$B=5/3$. The redshift scaling is more complicated: at $z \sim 0.6$, $d_A
\propto (1+z)^{1.3}$ leading to $C \sim -1.6$. 

For resolved detections, the cluster is spread over several resolution
elements:
the number of detector resolution elements subtended by
a cluster of size $R$ scales as $R^2/D_A^2$ and thus the integrated
noise $N_{int}$
scales as $M^{1/3}(1+z)^{-1}D_A(z)^{-1}$ (assuming white noise). 
If \nSN \ traces the integrated quantity $Y/N_{int}$, 
we obtain $\nSN \propto M^{4/3} (1+z)^2 d_A^{-1}$ and thus
would expect $B \sim 4/3$ %in Eq.~\ref{eqn:mass_scaling_app},
and $C \sim 0.7$ at $z\sim 0.6$. 

From these scaling arguments we would expect the slope parameter $B$
to reside within the range $1 \leq B \leq 1.66$, while $C$ is strongly
dependent on the experiment.
In practice, the true scaling is likely to be
influenced by a number of observational effects. For instance, small
clusters are partially washed out by the $1^\prime$ SPT beam. On the
other hand, large, low-redshift clusters are de-weighted by the matched
filter due to CMB confusion, reducing their detection significance
(the latter could be alleviated by multi-frequency cluster
detection).  

To better capture the possible form of the redshift evolution in the
SPT mass-significance scaling relation, a second parameter could be
incorporated (for example, an additional power-law dependence on the
angular diameter distance or linear evolution in the parameter $C$ with
redshift). Given the small size of the SPT sample and the
uncertainties placed on the other parameters, the
statistical power is insufficient to constrain such a parameter.
X-ray follow up observations of the SPT cluster sample are currently underway and will
shed much light on the scaling between SPT detection significance and SZ flux.

During the cosmological analysis (\S\ref{sec:cosmology}),
the parameters $A$, $B$, $C$, and the intrinsic scatter were
marginalized over with the Gaussian priors given in Table \ref{tab:marg_scaling_params}.
The preferred scaling relation parameter values for the highest likelihood
point in each of the four MCMC chains are given
in Table \ref{tab:marg_scaling_params}.
The recovered values and uncertainties on
$\sigma_8$ and $w$ are unaffected by widely varying priors on both
$B$ and $C$, indicating that these are not a significant source of
overall uncertainty in the current analysis.

\begin{table*}[!h]
\begin{minipage}{\textwidth}
\centering
\caption{Preferred Scaling Relation Parameters} \small
\begin{tabular}{l cc cccc}
\hline\hline
\rule[-2mm]{0mm}{6mm}
Chain						& $A$		& $B$		& $C$		& $Scatter$ 		\\
\hline
\LCDM \ WMAP7+SPT 			& 5.62	& 1.43	& 1.40	& 0.21	\\
\LCDM \ CMBall+SPT			& 5.46	& 1.43	& 1.38	& 0.21	\\
\wCDM \ WMAP7+SPT			& 6.01	& 1.42	& 1.45	& 0.21	\\
\wCDM \ WMAP7+BAO+SNe+SPT	& 5.42	& 1.44	& 1.41	& 0.21	\\
\\
(Gaussian Prior $\pm~ 1\sigma$)	& $6.01\pm1.8$	&$1.31\pm0.26$& $1.6\pm0.8$	& $0.21\pm0.04$	\\

\hline \\
\end{tabular}
\label{tab:marg_scaling_params}
\tablecomments{These are the values which maximize the likelihood for the highest likelihood point in each chain. For comparison, bottom row shows the Gaussian prior assumed on each parameter during the cosmological analysis.}
\end{minipage}
\end{table*}

\newpage
\section{Mass Estimates}
\label{app:mass_estim}

To provide a mass estimate for the catalog presented here, one must estimate $\nSN$. As mentioned in Section \ref{sec:scaling_relation}, $\SN$ is a biased estimator of $\left<\SN\right>$ and hence of $\nSN$. In order to correct for this bias, one must make an estimate of the mass function. Although the shape of the mass function is in principle dependent on the cosmology, we find that the mass estimates are robust to exact choice of cosmology and hence assume a fiducial cosmology for the mass estimates presented in Table \ref{tab:mass_estim}. Specifically, we use the maximum likelihood point in the $\Lambda$CDM WMAP7+SPT chain: ($\Omega_B h^2 = 0.0229$, $\Omega_C h^2 = 0.107$, $H_0 = 72.7$, $\tau = 0.0877$, $n_s(0.002) = 0.966$, $A_s(0.002) = 23.6 \times 10^{-10}$, $A=5.62$, $B = 1.43$, $C = 1.40$, $\mathrm{scatter} = 0.21$). The uncertainties in the cluster redshifts were assumed to be
negligible for these purposes.

Mass estimates were obtained for an assumed 
mass-\SN\ scaling relation using Bayes' theorem and assuming that the
prior probability for the mass is proportional to the \citet{tinker08} mass function:
\begin{equation}
{dP(\ln M|\SN) \over d \ln M} \propto { dN \over d\ln M} P(\SN|\ln M) \ .
\end{equation} 
The probability distribution $P(\SN|\ln M)$ was found from combining
the (logarithmic) intrinsic scatter in the scaling relation, 
as well as the (linear) scatter in $\SN$ for each cluster. 
This yielded an unbiased estimate of the mass of each cluster 
and provided the statistical uncertainty on each mass.

The uncertainty in mass estimates caused by uncertainty in 
the scaling relation was calculated
using a first order Taylor expansion about the best fit 
parameters $b_k^0$, where $b_k = \{A, B, C\}$. 
Changes in cluster mass estimates, $m_i$, caused by a different assumed
scaling relation are then given by 
$\delta m_i= \sum_k \frac{\partial m_i}{\partial b_k} (b_k-b_k^0)$. 
The covariance of the mass estimates is given by 
$\mathrm{cov}(\delta m_i,\delta m_j) = \sum_{k,l} \frac{\partial m_i}{\partial b_k} \sigma_{kl}^2 \frac{\partial m_j}{\partial b_l} $, where $\sigma_{kl}$ represents the covariance between parameters $b_k$ and $b_l$. This parameter covariance was found by numerically calculating the Hessian of the 10-dimensional log-likelihood surface about the best fit parameters.
In order to calculate the Hessian of the WMAP5 likelihood surface, we made use of CosmoMC package \citep{lewis02a}.
Assuming Gaussian errors, we inverted the Hessian matrix, yielding the parameter covariance. 
The systematic uncertainties in mass estimates reported in Table D1
are derived from the diagonal elements of $\mathrm{cov}(\delta m_i,\delta m_j)$. 

Uncertainty in the scaling relation parameters dominates the total uncertainty on the mass and the errors on the mass estimates are therefore 
strongly correlated. In Table \ref{tab:mass_estim}, the statistical and systematic errors are separated. The full covariance matrix is provided at http://pole.uchicago.edu/public/data/vanderlinde10/.

\begin{table*}[h]
\begin{minipage}{\textwidth}
  \centering
  \caption{Mass Estimates for the SPT Cluster Catalog} \small
  \begin{tabular}{l c c c c}
    \hline \hline
    \rule[-2mm]{0mm}{6mm}
    Object Name	& $\SN$ & $z$	& $M_{200}(\rho_{mean})(10^{14}\,\msun h^{-1})$ 
    						& $M_{500}(\rho_{crit}) (10^{14}\,\msun h^{-1})$ \\
&&	& $\mathrm{mass} \pm 68\% \mathrm{stat} \pm 68\% \mathrm{syst}$
	& $\mathrm{mass} \pm 68\% \mathrm{stat} \pm 68\% \mathrm{syst}$\\
    \hline
SPT-CL J0509-5342 &  6.61 &  0.4626 &   5.09  $\pm$   1.02  $\pm$   0.69  & 2.98  $\pm$   0.66  $\pm$   0.40  \\
SPT-CL J0511-5154 &  5.63 &  0.74 &   3.49  $\pm$   0.87  $\pm$   0.43  &   2.15 $\pm$   0.57  $\pm$   0.26  \\
SPT-CL J0516-5430 &  9.42 &  0.2952 &  7.84  $\pm$   1.29  $\pm$   1.25  & 4.39  $\pm$   0.84  $\pm$   0.69  \\
SPT-CL J0521-5104 &  5.45 &  0.72 &   3.39  $\pm$   0.89  $\pm$   0.39  &   2.08 $\pm$   0.58  $\pm$   0.23  \\
SPT-CL J0528-5259 &  5.45 &  0.7649 &   3.31  $\pm$   0.86  $\pm$   0.38  & 2.04  $\pm$   0.56  $\pm$   0.23  \\
SPT-CL J0533-5005 &  5.59 &  0.8811 &   3.19  $\pm$   0.79  $\pm$   0.37  & 1.99  $\pm$   0.53  $\pm$   0.22  \\
SPT-CL J0539-5744 &  5.12 &  0.77 &   3.05  $\pm$   0.84  $\pm$   0.33  &   1.89 $\pm$   0.55  $\pm$   0.20  \\
SPT-CL J0546-5345 &  7.69 &  1.0670 &   4.02  $\pm$   0.72  $\pm$   0.55  & 2.53  $\pm$   0.52  $\pm$   0.34  \\
SPT-CL J0551-5709 &  6.13 &  0.4231 &   4.84  $\pm$   1.06  $\pm$   0.68  & 2.81  $\pm$   0.67  $\pm$   0.39  \\
SPT-CL J0559-5249 &  9.28 &  0.6112 &   6.16  $\pm$   1.02  $\pm$   0.85  & 3.68  $\pm$   0.71  $\pm$   0.50  \\
SPT-CL J2259-5617 &  5.29 &  0.1528 &   4.91  $\pm$   1.45  $\pm$   0.71  & 2.68  $\pm$   0.82  $\pm$   0.38  \\
SPT-CL J2300-5331 &  5.29 &  0.29 &   4.37  $\pm$   1.27  $\pm$   0.60  &   2.48 $\pm$   0.75  $\pm$   0.33  \\
SPT-CL J2301-5546 &  5.19 &  0.79 &   3.07  $\pm$   0.84  $\pm$   0.30  &   1.90 $\pm$   0.55  $\pm$   0.18  \\
SPT-CL J2331-5051 &  8.04 &  0.5708 &   5.64  $\pm$   0.99  $\pm$   0.80  & 3.36  $\pm$   0.67  $\pm$   0.46  \\
SPT-CL J2332-5358 &  7.30 &  0.32 &   6.21  $\pm$   1.15  $\pm$   0.94  &   3.52 $\pm$   0.74  $\pm$   0.52  \\
SPT-CL J2337-5942 & 14.94 &  0.7815 &  7.86  $\pm$   1.17  $\pm$   1.18  & 4.77  $\pm$   0.86  $\pm$   0.70  \\
SPT-CL J2341-5119 &  9.65 &  0.9984 &   5.02  $\pm$   0.82  $\pm$   0.69  & 3.14  $\pm$   0.60  $\pm$   0.42  \\
SPT-CL J2342-5411 &  6.18 &  1.09 &   3.21  $\pm$   0.70  $\pm$   0.43  &   2.03 $\pm$   0.48  $\pm$   0.27  \\
SPT-CL J2355-5056 &  5.89 &  0.35 &   4.86  $\pm$   1.13  $\pm$   0.70  &   2.79 $\pm$   0.70  $\pm$   0.39  \\
SPT-CL J2359-5009 &  6.35 &  0.76 &   3.98  $\pm$   0.84  $\pm$   0.54  &   2.45 $\pm$   0.57  $\pm$   0.32  \\
SPT-CL J0000-5748 &  5.48 &  0.75 &   3.36  $\pm$   0.87  $\pm$   0.41  &   2.07 $\pm$   0.57  $\pm$   0.25  \\
    \hline
  \end{tabular}
  \label{tab:mass_estim}
  \tablecomments{Note that the masses $M_{500}(\rho_{crit})$ (where the overdensity is with respect to the critical density rather than the mean density) were calculated by converting from $M_{200}(\rho_{mean})$ assuming a Navarro-Frenk-White density profile and the mass-concentration relation of \citet{duffy08}. These masses may be underestimates for low-redshift clusters ($z\lesssim0.3$), where a power-law scaling relation fails to fully capture the behavior of the CMB-confused SZ signal, and that given for SPT-CL 2332-5358 should be considered a lower limit due to the known point source contamination.}
  \normalsize
\end{minipage}
\end{table*}

\end{document}